\documentclass[12pt,preprint]{aastex}
\slugcomment{Draft}

\shorttitle{Polar Regions of Cas A}
\shortauthors{Laming \& Hwang}

\begin{document}

\title{The Polar Regions of Cassiopeia A: The Aftermath of a
Gamma Ray Burst?}


\author{J. Martin Laming\altaffilmark{1}, Una Hwang\altaffilmark{2}, Balint Radics\altaffilmark{3},
Gergely Lekli\altaffilmark{3}, \& Endre Tak\'acs\altaffilmark{3}}


\altaffiltext{1}{E. O Hulburt Center for Space Research, Naval
Research Laboratory, Code 7674L, Washington DC 20375
\email{laming@nrl.navy.mil}}
\altaffiltext{2}{NASA/GSFC Code 662,
Greenbelt MD 20771,
also Department of Physics and Astronomy, Johns Hopkins University,
3400 Charles St, Baltimore MD 21218\email{hwang@orfeo.gsfc.nasa.gov}}
\altaffiltext{3}{Institute for Experimental Physics, University of Debrecen,
Bem t\'er 18/a, Hungary, H-4026}

\begin{abstract}
Probably not, but it is interesting nevertheless to investigate just
how close Cas A might have come to generating such an event.
Focusing on the northeast jet filaments, we analyze the polar
regions of the recently acquired very deep 1 Ms Chandra X-ray
observation. We infer that the so-called ``jet'' regions are indeed
due to jets emanating from the explosion center, and not due to
polar cavities in the circumstellar medium at the time of explosion.
We place limits on the equivalent isotropic explosion energy in the
polar regions (around $2.3 \times 10^{52}$ ergs), and the opening
angle of the x-ray emitting ejecta (around 7 degrees), which give a
total energy in the NE jet of order $10^{50}$ ergs; an order
of magnitude or more lower than inferred for ``typical'' GRBs. While the Cas
A progenitor and explosion exhibit many of the features associated
with GRB hosts, e.g. extensive presupernova mass loss and rotation,
and jets associated with the explosion, we speculate that the recoil
of the compact central object, with velocity 330 km s$^{-1}$, may
have rendered the jet unstable. In such cases the jet rapidly
becomes baryon loaded, if not truncated altogether. Although
unlikely to have produced a gamma ray burst, the jets in Cas A
suggest that such outflows may be common features of core-collapse
SNe.
\end{abstract}



\section{Introduction}
The speculation that gamma-ray bursts, or a subset thereof, might be
connected with core-collapse supernovae was initiated by the
coincidence of SN 1998bw with GRB 980425 \citep{galama98}, and has
been reinforced in recent years by the connections between SN 2001ke
and GRB 011121 \citep{garnavich03}, SN 2003dh and GRB 030329,
\citep{hjorth03}, and SN 2003lw and GRB 031203 \citep{malesani04}.
Spectroscopy of the afterglows of GRB 011211
\citep{reeves02,reeves03,rutledge03,butler05}, and of GRB 030227
\citep{watson03} suggest the presence of highly charged ions of Mg,
Si, S, Ar, and Ca, but not Fe, Co or Ni, (though Ni may be present
in GRB 011211). Claims of the detection of spectral lines in the
afterglows of GRB 001025A and GRB 010220 also exist
\citep{watson02}, though both these gamma-ray bursts lack definitive
SN connections. More recently though, \citet{sako05} have questioned
the statistical significance of these GRB afterglow line
identifications. It is against the background of these exciting
developments that we turn our initial attention in analysis of the 1
Ms Chandra observation of the Cassiopeia A supernova remnant to the
polar regions, with their striking ``jet-like'' morphologies. We are
motivated to examine in detail the spectra of ejecta knots in the
so-called jet regions with a view to determining whether these
regions are indeed due to an asymmetric explosion, and not just due
to cavities in an asymmetric circumstellar medium, and if due to
jets, whether or not we can infer energetics and other parameters
connected with the jet nature.

Ideas that core-collapse supernovae, particularly those occurring
after extensive presupernova mass loss as Type Ib/c, might be
inherently asymmetrical have also gathered force following SN 1987A
\citep[see e.g.][and references therein]{wang02}.  Linear
polarizations of optical light of order a few percent are typically
seen, suggesting nonspherical scattering stellar envelopes, with
aspect ratios as large as 2. Mechanisms by which core collapse
supernovae become asymmetrical generally derive from the rotation of
the progenitor, which as it collapses may or may not produce strong
magnetic field by a magneto-rotational instability.  A comprehensive
review of recent work in this area is given by \citet{wheeler04}.
Such ideas are appealing in the context of Cassiopeia A since
rotating core-collapse and the associated magnetic fields
\citep[e.g.][]{proga03,proga05,akiyama03} or anisotropic neutrino
emission \citep[e.g.][]{yamasaki05} are often suggested as a means
of producing bipolar outflows or jets. Before proceeding further in
this direction we need to be sure that the morphology we refer to as
``jets'' really are due to a feature of the explosion and are not
arising as a result of cavities in the circumstellar medium at the
appropriate locations. \citet{blondin96} model the latter situation,
and elongated structures of supernova ejecta resembling jets are
easily produced. Cas A is inferred to have undergone extensive mass
loss from its original 20-25 $M_{\sun}$ progenitor, to have been
only 3-4 $M_{\sun}$ upon explosion \citep{laming03,chevalier03}.
Additionally, the surrounding remnant stellar wind is relatively
dense and slow moving. Such conditions do not generally arise with a
radiatively driven wind from a 20-25 $M_{\sun}$ progenitor
\citep{woosley93}, but require the existence of a binary companion
to aid the mass loss. In such a case one might expect a departure
from spherical symmetry in the wind, from the binary orbital plane
to the rotation axis, as is apparent in the case of SN 1987A
\citep{sonneborn98}.

In this paper we use spectra from the newly acquired deep Chandra
observation to investigate the properties of the polar regions. We
describe the data and methods of analysis in section 2. In section 3
we demonstrate that models of the polar regions based on a
circumstellar cavity to explain the morphology cannot reproduce the
observed spectra, and that we really are seeing the evolution into
the remnant phase of a bipolar explosion. Section 4 discusses the
nature of these jets in more detail and section 5 concludes.

\section{Observations and Data Analysis}

We use spectra of the polar regions of Cas A from the Chandra
Observatory Very Large Project 1 Ms observation.  These data were
taken with the backside-illuminated S3 CCD chip of the Advanced CCD
Imaging Spectrometer (ACIS) in 9 observation segments, mostly over two
weeks in 2004, as described in more detail in Hwang et al. (2004).
The data were corrected for the time-dependent gain across the S3
chip, but could not be corrected for charge-transfer inefficiency
because the photon events were graded on board the spacecraft prior to
data telemetry.

Several regions along the three main filaments of the NE jet were
chosen for spectral analysis, as shown in Figure 1, as well as the
composite of the two main filaments of the much fainter counterjet in
the southwest (see Hwang et al. 2004).  The southernmost filament in
the NE is the straightest and longest of the three, and extends
farthest into the interior of the remnant. Of particular interest are
the very faint knots at the very outer tip of the NE jet filaments that
are newly revealed by the deep Chandra observation.  Their spectra are
shown in Figure 2.

All the spectra were extracted individually for each of the nine
observation segments, and corresponding detector response files
generated using CIAO v3.0.2 (http:www.cxc.harvard.edu/ciao).  The
individual spectra were then added together to produce the final
spectrum for a particular region, while the individual response files
were weighted according to the relative exposure time of each
observation segment before being added to produce the final response
files.  The time-dependent accumulation of soft X-ray absorbing
contaminant on the ACIS detector was modelled during spectral fitting
with the ACISABS model component.  The detector gain was adjusted to
optimize the fits, compensating in part for uncertainties in the
energy scale, due both to detector performance and the intrinsic bulk
motion of the gas.

The goal of the spectral analysis is to obtain electron temperature,
ionization age (the integral of electron density over time since shock passage
for the reverse shocked ejecta, assuming a constant electron temperature
during this time), and
element abundances.  We follow Laming \& Hwang (2003) in fitting the
spectra with models for single-temperature plasmas with time-dependent
ionization (nonequilibrium ionization or NEI).  We also experimented
with collisional ionization models, NEI models with multiple
components, and NEI models with a range of ionization ages (mimicking
plane-parallel shocks).  For the knots along the main body of the jet
filaments, the spectral fits were generally improved by using a
plane-parallel shock (pshock) model compared to a single NEI
component.  Two-component NEI models were also an improvement, but
generally did not perform better than the pshock models.  It is
noteworthy that the knots at the very tip of each jet filament have
spectra that are well-described with a single component (although this
is at least partly because these spectra contain fewer counts).  The
ionization ages fitted for the jet tip spectra are very high, and they
may also be successfully fitted with collisional ionization
equilibrium models (see Table 1).  All the jet region spectra show
strong emission lines of Si, and S.  Those with sufficient counts also
show Ar and Ca as well as Fe K emission.  Except for the rather
extended knot at the tip of the middle NE filament, the jet tip
spectra are effectively cut off at around 5 keV.

There are various possibilities as to which elements contribute to the
continuum.  Following Laming \& Hwang (2003), we generally model the
continuum as coming from either ionized O or Si, with an O continuum
consistently giving a better fit (those results are given in Tables 1
and 3).  On the basis of $\chi^2$ alone, however, it is not possible
to confidently distinguish between the various continuum models. The
assumption of which light elements provide the continuum does not
strongly affect the quality of the spectral fits but does affect
inferences concerning the electron densities and masses in the knot.
H continuum models, for example, require higher densities and masses
than those with O continuum.

Given the complexity of the spectral models required for most of the
jet knots, it is helpful to examine properties of the line emission to
assess qualitative trends in the data.  In particular we are
interested in possible trends that might be present with location
along the jet in the line ratios or line equivalent widths.  We
therefore fitted the spectra with line blends of individual elements
and a bremsstrahlung continuum, as summarized in Table 2.  We
illustrate in Figure 3 the two strongest trends that we found.  The
first panel shows Fe K EQW plotted against the projected distance of
the jet region from the point source near the center of the remnant.
With the exception of the knots at the very tip of the jet (where the
temperatures are significantly lower than elsewhere along the jet,
limiting both the emissivity of and sensitivity to Fe K), the Fe K
line strength is seen to either stay roughly level or increase with
distance outward along the jet filament.  The northern filament in
particular appears to show the strongest tendency for an outward
increase in Fe K line strength.  Although the line strength of Fe K is
affected by a number of factors, this trend appears to be echoed in
the fitted Fe abundances as well.  The second panel of the figure
shows the correlation between Ca He $\alpha$ EQW and Fe K EQW.  The Ca
He $\alpha$ blends in the jet knots are unusually prominent compared
to other regions of the remnant (Hwang et al. 2000), and are seen to
roughly follow the Fe K line strength along the jet filaments.

In the jet models discussed below, a significant fraction of reverse
shocked jet plasma has now cooled by radiation and adiabatic expansion
to temperatures such that it would no longer emit X-rays. The
extrapolation of the observed equivalent width trends to larger
distances outward along the jet is consistent with such models. If the
element composition of the presently unseen jet tip material were
dominated by Fe, as would be consistent with its origin deep inside
the progenitor, then the radiative cooling time of such plasma would
be even shorter than for the O dominated composition assumed so far,
allowing it to cool very quickly.

\section{Circumstellar Cavities or Ejecta Jets?}
\subsection{Circumstellar Cavities}
The bi-polar morphology exhibited in Cas A has two plausible
origins. Both a symmetric explosion into an asymmetric circumstellar
medium, or an asymmetric explosion into a symmetric medium could
produce jet-like structures. The first possibility has been considered
by \citet{blondin96}. In core-collapse supernovae, asymmetries in the
circumstellar density may arise from non spherically symmetric
pre-supernova mass loss, as has been established in the case of SN
1987A. The high degree of mass loss in Cas A inferred from the
dynamics \citep{laming03}, and the likelihood that such winds
were driven by the interaction of the progenitor with a binary
companion lend support to this idea.

We simulate the evolution of the ionization balance for a knot with
composition resembling that in the middle jet tip knot discussed
above, i.e O:Si:Fe being 0.82:0.13:0.05 by mass. The departure from
spherical symmetry means that forward shocked circumstellar plasma
at the head of the jet moves non-radially (i.e. tangentially) away
from the apex, with the consequence that the contact discontinuity
at the apex is now much closer to the forward shock. Hence the
forward shock in the polar region should be close to the outermost
X-ray emitting ejecta, at about 3.8 pc, giving an aspect ratio of
3.8/2.5 = 1.5, taking the equatorial forward shock radius as 2.5 pc.
Such an aspect ratio agrees best with a density contrast of a factor
of eight between equatorial and polar regions \citep[see Figs 3 and
4 in][]{blondin96}. Accordingly we take our equatorial core-envelope
model from \citet{laming03}, with an outer envelope power law
density profile with an exponent of 9 and a uniform density inner
core, and reduce the circumstellar density by a factor of 8 in our
initial study. By following the evolution of the ionization balance
and electron and ion temperatures in the ejecta following reverse
shock passage, we compute, and plot in Figure 4, the locus of
electron temperature against ionization age for this model, as well
as for models with outer envelope power law exponents of $n=7$, and
11. The upper branch of each curve corresponds to the phase when the
reverse shock is propagating through the constant density inner
ejecta core. The lower branch corresponds to reverse shock
propagation through the outer power law envelope, and the maximum
ionization age is found at the uppermost point in each curve at the
core-envelope boundary. For $n=9$, this is found to be $4.3\times
10^{11}$ cm$^{-3}$s ($\log n_et = 11.6$), with a temperature of
$6.5\times 10^6$ K ($\log T_e = 6.8$). The other models either give
higher ionization age and lower temperature, or the reverse.
Comparison with the fit results from section 2, plotted as boxes
corresponding to the uncertainties in temperature and ionization age
for each knot, shows that no cavity model gives sufficient density
of plasma at high enough temperature to match the observations.
Simply put, the ejecta expand sufficiently rapidly into the cavity
that the density is too low either for appreciable electron-ion
equilibration to raise the electron temperature or to ionize the
plasma to the observed values. The only way to make such a model
work would be to include some collisionless electron heating at the
reverse shock \citep[currently neglected, see][and and references
therein for fuller discussion of this point]{laming03}, which with
reference to Figure 4 could provide the extra factor of three or so
in electron temperature required to match the observations of knots
in the jet ``stem'', but those at the jet tip would still be
discrepant.

\subsection{Jets}
We model jets based on the simulation of \citet{khokhlov99}. These
authors studied the explosion of the inner 4.1 $M_{\sun}$ of a 15
$M_{\sun}$ progenitor (assuming substantial presupernova mass loss),
induced by baryonic jets emanating from the central regions of the
star, where the composition is dominated by Fe and Si. The inner 1.6
$M_{\sun}$ is assumed to fall back onto a protoneutron star, and the
remaining 2.5 $M_{\sun}$ of ejecta should give a reasonable match to
the parameters of Cas A. Initially, each jet comprises about 0.05
$M_{\sun}$ of material and together they have $9\times 10^{50}$ ergs
of kinetic energy. The jet power is constant for the first 0.5 s,
and ramped down to zero after another 1.5 s.  The equivalent
isotropic mass and energy in the jet when they are launched would be
2 $M_{\sun}$ and $2\times 10^{52}$ ergs respectively, increasing to
8 $M_{\sun}$ and $8\times 10^{52}$ ergs upon jet breakout. From the
radius of the jet at launch compared to its final radius upon
emergence through the stellar surface, we infer a jet density
profile $\rho _j\propto 1/r$, which also matches the figures
presented by \citet{khokhlov99}. We use an adaptation of the
BLASPHEMER code \citep{laming02,laming03} which implements the
solutions for ejecta profiles $\rho\propto r^{-n}$ with $n<3$ given
by \citet{truelove99}, and is summarized here in Appendix A. A jet
unconfined by the surrounding ejecta would obey $\rho _j\propto
1/r^2$.

In Figure 5 we plot the locus of electron temperature and ionization
age for jet models assuming $2.3\times 10^{52}$ equivalent isotropic
energy and 1.815 $M_{\sun}$ ejecta mass. The ejecta mass is taken
from that determined in the polar region exterior to the jet by
\citet{laming03}. The energy is modified slightly from the value
used by \citet{khokhlov99} to give better agreement with the
observed morphology. Specifically, we want the faint X-ray knots
observed at the very tip of the jet to end up at a radius of
approximately 3.8 pc from the explosion center. The models differ in
that abundance sets of O:Si:Fe are taken from the fits to the three
jet tip knots, color coded as black - north tip, 0.004:0.628:0.368;
red - middle tip, 0.82:0.13:0.05; and green - south tip,
0.55:0.12:0.33. Additionally we have also computed a model
appropriate for the north tip, where half the plasma is assumed to
be H. This is given by the black dotted line. As can be seen, the
jet models give a reasonable match to both the CIE knots at the jet
tip, as well as the NEI knots further down the jet stem, in terms of
ionization age and electron temperature. Some of the jet stem knots
fall at lower temperatures than modeled assuming O to be the most
abundant element, but which could be understood if the jet knot
plasma comprises a substantial fraction of H. Of course the
abundances in the jet stem knots are probably not the same as those
in the three tip knots, although here they have been modeled as
such. The variation between the three curves for N, M, and S jet
filaments should give some idea as to the variation to be expected
as abundances vary within reasonable limits.

As well as comparing electron temperatures and ionization ages
coming from our models with those from fits, we have also
investigated the ionization balance predicted by the models for Fe
and Si in the jet tip knots. Figure 6 shows the model Fe and Si
charge state distributions for ejecta that went through the reverse
shock at various times between 1.3 and 1.6 years after explosion,
compared with the charge state distribution of a plasma in
collisional ionization equilibrium at $8.6\times 10^6$ K (the thin
solid line), meant to match the conditions in the jet tip knot of
the middle filament. For both Fe and Si, the model charge states
corresponding to an electron temperature of $8.6\times 10^6$ K are a
poor match to the observed charge states, with too many ions in high
charge states, e.g. Fe$^{24+}$ and Si$^{13+}$. This is because the
recombination from these ions is a slow process. Much better
correspondence between observation and the models is found for
ejecta that have been allowed to cool down to temperatures $5 - 6
\times 10^6$ K, i.e. lower than actually observed. We speculate that
although we have tried to model and fit the knot spectra as though
they are single plasmas, shocked instantaneously by the reverse
shock, some inhomogeneity may still exist. Small variations in
elemental composition or density are likely to reveal themselves
during the onset of thermal instability, when the temperature is
decreasing very quickly with time, making the heavy element line
emitting regions cooler than the surroundings which mainly emit
continuum. At 1-2 years after explosion, the reverse shock speed in
our models is 4000-5000 km s$^{-1}$, which means it would take
$10^8$ seconds or 3 years to traverse a knot of 1 arcsecond extent.
It is therefore also likely that our assumption that the whole knot
is shocked instantaneously breaks down, in that the reverse shock
speed evolves as it traverses the knot making a single ionization
age less realistic.

A related problem is that knots in the main body of the jet are
better fit by plane-parallel shock models that integrate over a
range of ionization age, rather than by single ionization age
models. These knots with smaller ionization ages are shocked later
(i.e. around 5 - 10 years after explosion) than those at the jet
tip, at a time when the reverse shock in our models is evolving much
less rapidly with time. These knots are also observed at smaller
radii relative to those in the jet tip than our models would
predict, and have similar problems with the ratios of He-like to
H-like Si, in that the models predict much more H-like than actually
observed. This is quite different from Si in knots of similar
temperature and ionization age studied in \citet{laming03}. The
difference lies in the fact that for the jet models considered here,
the electron temperature was previously much higher than it is at
present, and the plasma is recombining. The NEI models which assume
a constant temperature are probably not good approximations. This is
quite different to the knots in more equatorial regions in
\citet{laming03}, where the postshock knot electron temperature is
roughly constant, making the NEI models excellent approximations. We
suspect that these knots may have initially undergone an oblique
interaction with the reverse shock, leading to a lower peak electron
temperatures, and have undergone further interactions with shocks
reflected from the blast wave as it encounters clumps (i.e.
quasi-stationary flocculi) in the circumstellar medium, which would
further decelerate and heat them. Such a scenario does indeed seem
to be the case for SN 1987A \citep{zhekov05}, in that a variety of
shock conditions are required to fit the Chandra/LETG spectra, and
indeed should be expected as the blast wave begins to encounter the
inner ring of circumstellar material, producing a succession of
reflected shocks back into the following ejecta.

\subsection{Related Issues}
Our conclusion that we are indeed seeing an asymmetric explosion
into a symmetric medium is based solely on the spectroscopy of knots
in the polar regions. \citet{laming03} using similar techniques for
ejecta knots observed elsewhere in the remnant concluded that the
explosion energy per unit solid angle was higher in regions close to
the jet than in more equatorial directions. A completely independent
conclusion of similar anisotropy in the explosion comes from a
survey of the outlying optical knots by \citet{fesen01}. These
optical knots are generally taken to be essentially undecelerated by
their interaction with the circumstellar plasma, and hence any
velocity variation exhibited by these knots must have an origin in
the explosion itself, so that our conclusion is in line with those
of other researchers.

However considerable uncertainty surrounds the mechanism(s) of
formation of the various kinds of optical knots, so here we briefly
offer some speculations as to how these features fit in with the
hydrodynamic models for Cas A studied in \citet{laming03}. Knots
rich in nitrogen are observed around the remnant (but not apparently
in the NE jet region) expanding with velocity about 10,000 km
s$^{-1}$. Due to their composition, they must originate in the outer
layers of ejecta. We suggest that these knots are the end result of
a Rayleigh-Taylor instability at the contact discontinuity. In
simulations, R-T fingers of ejecta typically do not penetrate
through the forward shock, but \citet{jun96} show that in the
presence of a clumpy external medium, this is not the case. The R-T
fingers can gain extra kinetic energy from the vortices generated by
the shock-cloud interactions. In this case we would expect the
velocity with which the resulting clumps of nitrogen rich material
expand to be similar to the ejecta expansion velocity at the
core-envelope boundary, which in the models of \citet{laming03} is
indeed around 10,000 km s$^{-1}$. Knots rich in oxygen and its
burning products (they also show prominent S II emission) are found
in various regions of the remnant, especially in the NE jet. Many
authors have surmised that these arise as a result of the ``Fe
bubble'' effect, whereby Ni-Co-Fe bubbles inflated by the energy
deposition due to the radioactive decay of Ni to Co and ultimately
to Fe compress the surrounding plasma. This argument would suggest
that Fe should be present, or should have been present in the jet
regions in order to produce the fast moving knots observed there
today. We will return to the subject of Fe in the jets below.

Finally, the location and velocity of the reverse shock in the NW
and SW regions of the remnant has been inferred from optical
observations with HST WFPC2 images taken in 2000 and 2002
\citep{morse04}. The reverse shock radius of $\sim 1.8$ pc and
velocity with respect to the freely expanding ejecta of $\sim 2000$
km s$^{-1}$ correspond reasonably well with the $n=9$ model from
\citet{laming03}, which predicts a reverse shock radius of 1.7 pc
and velocity of 1500 km s$^{-1}$. The ejecta free expansion velocity
in this model is 4800 km s$^{-1}$, to be compared with 5000 km
s$^{-1}$ in \citet{morse04}. Adjusting the model of \citet{laming03}
to better match say the faster reverse shock velocity (e.g. by
decreasing the ejecta envelope power law index to $n=7$) would make
the discrepancy between the measured and modeled reverse shock radii
worse, and vice versa. We would expect the models to slightly
underestimate the reverse shock velocity at 340 years since
\citet{laming03} make the approximation of holding the reverse shock
velocity constant during its propagation through the uniform density
ejecta core, whereas in fact we would expect it to slowly
accelerate. Consequently the reverse shock velocities computed at
earlier times in the remnant evolution for a given model should be
more accurate than those for the current epoch. In any case, the
reverse shock is observed to be fragmented and irregular in
morphology due to local density inhomogeneities, and may even be
dynamically unstable, so we consider the agreement between our model
and the HST observations to be satisfactory.

\section{Discussion}
\subsection{Jet, Mass, Energy and Opening Angle}
We first discuss in more detail the nature of the \citet{truelove99}
$n=1$ models employed to simulate the jet emission. The equivalent
isotropic energy of $2.3\times 10^{52}$ ergs is chosen to place
ejecta cooling through temperatures of 0.7 keV at a distance of
about 3.8 pc from the explosion center. These models predict a blast
wave radius at the jet tip of 5.66 pc and a velocity of 7850 km
s$^{-1}$.  For reasons to be discussed below, these are probably
overestimates, but we do not actually detect the location of the
blast wave anywhere in the jet region.  This radius however is not
too far off from the radius of the outermost optical knot in the jet
of 4.8 pc \citep{fesen01}. The gas that we see today at 0.7 keV
temperature encountered the reverse shock between one and two years
after explosion and has a mass coordinate of about 0.5.  This means
that nearly half the ejecta went through the reverse shock during
these first one to two years, and has by now cooled to temperatures
below those where it would be visible in X-rays to Chandra.
Consequently in the jet there should be a lot of cold plasma at the
jet head, cooled by both radiative losses and adiabatic expansion.

To our knowledge, there are no published simulations of jet
explosions carried through to the supernova remnant phase of the
evolution. However we can get some idea of the hydrodynamics
involved in the interaction of a jet with an ambient medium from
recent laboratory work \citep{blue05,foster05,lebedev05}. Here we
can clearly see a jet stem, such as we plausibly observe in X-rays
in Cas A, and a denser ``mushroom cap'' jet tip, which in Cas A has
presumably cooled to too low a temperature to be visible in X-rays.
From Figure 5 it appears that the knots at the jet tip are on the
verge of becoming radiatively unstable. Radiatively cooled ejecta at
the jet tip will reduce the pressure driving the forward shock, so
that the forward shock radius and velocity might be expected to be
smaller than in the adiabatic models of \citet{truelove99}.

As mentioned above, we do not detect the blast wave in the NE jet region.
Elsewhere in the remnant it is easily identifiable as a rim of continuum
emission, presumably synchrotron emission from cosmic ray electrons.
The width of this rim is identified as being due to the radiative loss time of
the cosmic ray electrons as they are advected away from the shock
\citep{vink03}, with the result that the magnetic field must be of order
0.1 mG, presumably amplified from a very low value in the presupernova stellar
wind by a cosmic ray precursor as suggested by \citet{bell01} and
\citet{lucek01}. Why is
such emission absent from the jet region? One possibility is that the shock
here is oblique; the inflowing and outgoing plasma do not flow along the shock
normal, and diffusive shock acceleration is known to be less efficient in
such cases \citep[e.g][]{drury83}, leading to lower cosmic ray energies and
weaker magnetic field amplification by the cosmic ray precursor. It is also
possible that the blast wave at the jet tip is actually outside the Chandra
field of view.

Our simulations of the knot spectra above give the equivalent isotropic energy
in the jet as $2.3\times 10^{52}$ ergs, with the assumption that the jet
equivalent isotropic mass is 1.815 $M_{\sun}$, as determined from knots at
the jet base by \citet{laming03}. We are however, unable to independently
constrain the jet mass in the same way since as mentioned above, we do not
detect the blast wave driven by the jet. The plot of electron temperature
against ionization age is essentially independent of the jet mass and energy,
so long as these are constrained to place plasma cooling through a temperature
of about 700 eV at a radial distance from the explosion center of about 3.8 pc.
A higher jet mass \citep[as in e.g.][]{khokhlov99} would of course require a
higher jet energy, but we consider this an unlikely possibility. We have
made estimates of the plasma density and mass in the knots at the jet tip from the spectral fits,
which generally come out {\em lower} than our model values. While there is
considerable uncertainty in these estimates, they suggest that if anything the
jet has density similar to or less than the surrounding stellar envelope, and is not
overdense as in the case modeled by \citet{khokhlov99}.

Given an equivalent isotropic energy in the jet of $2.3\times
10^{52}$ ergs, the evaluation of the total jet energy requires some
knowledge of the opening angle. In Figure 7 we overlay blast wave
profiles including the NE jet estimated from a ``pseudo'' Kompaneets
approximation outlined in Appendix B. We assume that sufficient cold
ejecta, either shocked or unshocked, exists in the jet region so
that the remnant retains some ``memory'' of the initial energy
distribution. We take the jet energy profile to be constant within
the jet opening angle, and to fall off as $E\propto\theta ^{-n}$
outside this region. We consider values of $n$ of 2, 3, and 8,
coming from values favored by \citet{lazzati05}, \citet{zhang04} and
\citet{graziani05} respectively. The panels show the cases of jet
opening angles of 5 degrees (top left), 7 degrees (top right), 9
degrees (bottom left) and 11 degrees (bottom right), and in each
case the $n=8$ model is the narrowest and $n=2$ the widest. The
total jet energy evaluates to about $10^{50}$ ergs, taking an
opening angle of 7 degrees. This is nearly an order of magnitude
lower than the ``standard'' gamma-ray burst radiated energy found by
\citet{frail01}, and probably much lower than the kinetic energy of
relativistic ejecta in such events, and argues against Cas A
actually having produced a gamma-ray burst during its explosion.
This energy, with the likely jet underdensity compared to the
surrounding stellar envelope also suggest that the jets are unlikely
to have induced the explosion, being significantly less energetic
than those in the simulation of \citet{khokhlov99}. More likely they
are the by-product of an explosion proceeding by different means.

Our estimation of blast wave profiles around the jet region assumed
that at all times in the evolution of Cas A there has been
sufficient cold ejecta at the jet head to retain some ``memory'' of
the initial explosive energy distribution with angle. In section 2
we described how the equivalent width of the Fe K feature is seen to
increase with radial distance, especially in the northern filament,
a trend which if extrapolated would suggest that the jet tip regions
are dominated in composition by Fe. This would shorten the radiative
cooling time for shocked ejecta over that calculated assuming O
dominated composition, and support our assumptions here.
\citet{hwang03} estimate that only a few per cent of the total mass
of Fe that Cas A is expected to have ejected are currently visible
as X-ray emitting ejecta, and so a significant portion of the unseen
jet material could also be composed of Fe. A number of authors
\citep{wang02,nagataki98,nagataki03} have also suggested that
bipolar jets should be locations where $\alpha$-rich freeze out
occurs with correspondingly large abundances of $^{44}$Ti. In the
X-ray emitting portion of the NE jet, we see rather little Fe, and
certainly no knots that may be considered as pure Fe as found by
\citet{hwang03}. We have searched for inner shell line emission from
$^{44}$Sc and $^{44}$Ca (the decay products of $^{44}$Ti, formed
mainly by K shell electron capture) in the region at the head of the
jet where we expect the X-ray dark ejecta to be, without success.
Our (3 sigma) upper limit is about 4 photons, less than 0.1\% of the
$^{44}$Sc inner shell line emission expected to be observed in the
whole remnant, based on the gamma-ray line observations of nuclear
deexcitation in $^{44}$Sc \citep{vink01,vink03} and $^{44}$Ca
\citep{iyudin94,schonfelder00}.\footnote{A detected gamma-ray flux
of $2-3\times 10^{-5}$ photons cm$^{-2}$ s$^{-1}$ gives 20-30
photons cm$^{-2}$ in $10^6$ seconds, which combined with the
ACIS/HMRA effective area gives total detected photons in the range
$5\times 10^3 - 10^4$ in the remnant as a whole.} We should caution
that the regions we investigated are right at the edge of the
Chandra/ACIS field of view, and that regions that underwent
$\alpha$-rich freeze out may even be further out.

\subsection{Could Cas A Have Produced a GRB?}
As mentioned above, although Cas A was produced in an explosion that accelerated
jets of material, the effects of which persist to this day, the energetics of
these features make it unlikely that Cas A actually produced a gamma-ray burst.
In this section we discuss how other features of the Cas A SNR might reinforce
this conclusion.

The Cas A explosion energy, in the range $2 - 4 \times 10^{51}$,
ergs places it clearly as a supernova and not a hypernova, though an
observer placed on the jet axis (equivalent isotropic energy of
$2.3\times 10^{52}$ ergs) might have perceived it as such from line
broadening \citep{mazzali02,ramirez04}. \citet{wang99} identified possible
hypernova remnants in M101, supporting the suggestion
\citep{paczynski98,macfayden99} that such a class of highly
energetic explosions exists. Further, \citet{atoyan05} identify a source detected
only in TeV gamma-rays by HESS (High Energy Stereoscopic System) as the
remnant of a gamma-ray burst, based on theoretical arguments.
As \citet{ramirez04} discuss, the
absence of a Compton scattered jet component (i.e. scattering
photons from a misaligned jet into an observer's line of sight) in
otherwise normal Type Ibc SNe would suggest that those hosting GRBs
are not normal Type Ibc SNe, supporting the conjecture that Type Ibc
SNe hosting gamma ray bursts constitute a separate class of more
energetic SNe, known as hypernovae. However more recently, \citet{soderberg06}
present evidence that the optical properties of Type Ibc SNe associated with
GRBs are not significantly different to ``normal'' Type Ibc events, and
that GRB associated SNe can only be reliably identified from their radio
signatures. \citet{nagataki06} show that $^{56}$Ni production in a
GRB jet would be insufficient to power a hypernova, and that if these explosions
exist as a separate class, $^{56}$Ni must be produced elsewhere in the ejecta,
further loosening any association between hypernovae and jets.

\citet{podsiadlowski04} and \citet{vanputten04} estimate the
branching ratio from Type Ibc supernovae to gamma-ray bursts to be
in the range $10^{-3} - 10^{-2}$. \citet{podsiadlowski04} suggest
that special evolutionary circumstances are required to lead to an
explosion of the hypernova type and the accompanying gamma-ray
burst, presumably involving a specific type of binary interaction to
give the required rotation rate and mass loss. \cite{vanputten04}
identifies the small branching ratio with the probability of the
central compact object remaining at the explosion center and
accreting material to become a high-mass black hole. This argument
would suggest that Cas A, which produced what is presumably a
neutron star recoiling with velocity in the plane of the sky of 330
km s$^{-1}$ \citep{thorstensen01}, could not have produced a
gamma-ray burst. Aside from these statistical arguments, the
recoiling neutron star should naturally be expected to destabilize
the jets, causing rapid baryon loading and quenching any nascent
gamma-ray burst.

We have verified that the explosion center determined from the
positions and velocities of the X-ray knots cataloged by
\citet{delaney04} is consistent with that determined from optical
observations by \citet{thorstensen01}, although the uncertainties in
the X-ray position are considerably larger than those in the optical
work. In this case the central compact object recoil is at about 75
degrees to the jet axis. According to \citet{wang05}, the recoil
component perpendicular to the jet axis requires either a relatively
slow rotation rate upon explosion, rather unlikely in our view since
such a case appears unlikely to generate jets, or the presence of a
binary companion. A binary companion is also favored to produce the
required degree of pre supernova mass loss \citep{young05}, and so this is the
scenario we prefer.

Additional circumstantial evidence against Cas A hosting a gamma-ray burst
lies in the apparent absence of stellar wind circumstellar media in most
gamma-ray burst afterglows modeled to date \citep{chevalier00,panaitescu02},
with \citet{piro05} identifying one of the few (GRB 011121) consistent with
a stellar wind density profile. Of course Cas A shows clear evidence of
expanding into a remnant stellar wind, consistent with only a minority of
gamma-ray bursts. The Cas A explosion
also seems to have been extraordinarily dim, possibly due to being shrouded
in a dense stellar wind at the time of explosion, though \citet{young05}
comment that the dust component of the wind, if the Cas A progenitor evolved
as a Wolf-Rayet star, is unlikely to have significantly changed the interstellar
extinction from values found today.

\citet{mazzali05} report
the observation of the Type Ic supernova SN 2003jd that appears to be an
intermediate case between normal Type Ic SNe and gamma-ray burst host SNe.
No gamma-ray burst is noted, but the supernova is clearly asymmetrical,
exhibiting double peaked [O I] lines. These are interpreted as coming from
equatorial regions of the progenitors, while emission from the polar regions
is dominated by jets. More recently, \citet{folatelli05} and \citet{tominaga05}
report similar conclusions for SN 2005bf, another energetic Type Ib/c supernova.
We suggest that the explosion of Cas A may have been a
similar event. It is clearly asymmetrical, with a little more mass in equatorial
than in polar regions
\citep[taking the estimates of Table 5 in][at face value]{laming03}, presumably
the result of polar jets forcing the overlying stellar envelope to lower
latitudes. However the degree of asymmetry produced appears to be at the low
end of that predicted by models, and the jets themselves are clearly less
energetic than in either the simulations of \citet{khokhlov99} or in
observations of gamma-ray bursts \citep{frail01}.

\section{Conclusions}
We have studied the NE polar region of the Cas A SNR in considerable detail.
In contrast, the counter jet region appears to be expanding into a more complex
circumstellar medium, possibly a cloud of higher density, which limits the
conclusions we may draw. Fits to knots in the counter jet are completely
consistent with those in the NE jet stem. However the high ionization age knots
at the tip of the NE jet, which turn out to be of crucial importance for the
analysis in this paper have no counterparts in the counter jet. It is also worth
commenting that the three jet tip knots studied in detail here are simply not
visible on images from the earlier 50 ks Chandra observations of Cas A. The deep
1 Ms VLP observation was absolutely necessary to see these at all. Analyses of
the knots reveals that
the ``jet''-like morphology really is due to an explosive jet, and not arising
as a result of interaction with a cavity or lower density region in the
circumstellar medium. While Cas A exhibits many of the properties suggested
for gamma-ray
burst hosts; extensive presupernova mass loss and jets presumably requiring
nonnegligible rotation of the progenitor, the jet itself is inferred to be
significantly less energetic than generally accepted for gamma-ray bursts,
and so is unlikely to have actually been a gamma-ray burst itself. Any
pair or Poynting flux dominated jet is likely to have become baryon loaded
if the jet was destabilized by the recoil of the neutron star, giving rise to
the ejecta dominated structure that we see today. Further, the jet is also
significantly less energetic than in the jet-induced explosion model of
\citet{khokhlov99}, suggesting that the supernova giving rise to the Cas A
remnant was not a jet-induced explosion. This is supported by \citet{laming03}
who find evidence of asymmetry in the explosion at the low end of or below
the range generally modeled or invoked to explain polarization observations
of Type Ibc or  Type II supernovae. Consequently we believe we are seeing jets
produced as a by-product of an explosion that proceeded by other means, e.g. a
more usual neutrino generated event, and that evidence of convective overturn
in the ejecta should be interpreted in terms of these types of models
\citep[e.g.]{kifonidis00,kifonidis03} rather
than the rotating jet powered models \citep[e.g]{akiyama03}.

This work only scratches the surface of the analysis to be done and the physics
to be extracted from the Cas A VLP dataset. In subsequent papers we intend to
analyze in detail the ejecta knots, particular emphasis on the Fe rich regions.
Using methods developed in
\citet{laming03} and \citet{hwang03}, together with observed knot velocities,
both from proper motions in the plane of the sky and from Doppler shifts, to
reconstruct a three dimensional distribution of ejecta knots. With this in hand,
we then envisage studying various explosion models to gain insight into the
mechanisms and instabilities involved by comparison with our catalogue of knots.

\acknowledgements
This work was supported by grants from the CXO GO and the NASA LTSA
programs. JML was also supported by basic research funds of the Office of
Naval Research.

\appendix

\section{Jet Models in BLASPHEMER}
In \citet{laming03} and \citet{hwang03} we used core-envelope solutions given by
\citet{truelove99} for trajectories of the forward and reverse shocks, where the
ejecta density distribution is taken to be represented by a uniform density core
with an outer power law envelope $\rho\propto r^{-n}$, where $n>5$. Here we outline
our implementation of the \citet{truelove99} solutions for power law ejecta
density distributions, $\rho\propto r^{-n}$ with $n<3$, which have no constant
density core, expanding into a circumstellar medium with $\rho\propto r^{-s}$.
For reasons given in the main text, we expect these to be better
representations of the jet regions, taking $s=2$ as in \citet{laming03}.
We use the same system of units as in
\citet{laming03}\footnote{The numerical factors in equations A1 and A2 in
\citet{laming03} for the $s=0$ ambient medium are in error; they should be
473.6 and 3.43, instead of 423 and 3.07, respectively.}. Other parameters are;
the lead factor $l_{\rm ED}=1.1+0.4/\left(4-s\right)$, the ratio of forward
reverse postshock pressures, $\phi _{\rm ED}=0.343\left(1-s/3\right)^{0.43}$,
and the expansion velocity of the outermost ejecta
$v_{\rm ej}=2\sqrt{\left(5-n\right)/\left(3-n\right)}$.

The blast wave trajectory for $n<3$ ejecta during the ejecta dominated phase
is given by
\begin{equation}
t={R_b\over v_{ej}l_{\rm ED}}
\left[1-\left(3-n\over 3-s\right)\sqrt{\phi _{\rm ED}\over
l_{\rm ED}f_n}R_b^{\left(3-s\right)/2}\right]^{-2/\left(3-n\right)}.
\end{equation}
We define a transition time to Sedov-Taylor behavior as follows. The blast wave
radius at the transition is estimated from
$v_b=R_b/t=2\sqrt{\xi}R_b^{\left(s-3\right)/2}
/\left(5-s\right)$ with $t=R_b/v_{\rm ej}l_{ED}$ yielding
$R_{\rm conn}=\left(v_{\rm ej}l_{\rm ED}\left(5-s\right)/
2\sqrt{\xi}\right)^{2/\left(s-3\right)}$. The transition time is then calculated
from equation A1 with $R_{\rm conn}$, where $\xi =\sqrt{\left(5-s\right)\left(10
-3s\right)/8\pi }$. For $t<t_{\rm conn}$, $R_b$ is calculated from
equation A1, $R_r=R_b/l_{\rm ED}$,
\begin{equation}
v_b={R_b\over t}\left\{1+{n-3\over 3-s}\sqrt{\phi _{\rm ED}\over l_{\rm ED}f_n}
R_b^{\left(3-s\right)/2}\over 1+{n-s\over 3-s}\sqrt{\phi _{\rm ED}\over
l_{\rm ED}f_n}
R_b^{\left(3-s\right)/2}\right\},
\end{equation}
and $v_r=R_r/t-v_b/l_{\rm ED}$. For $t>t_{\rm conn}$,
\begin{eqnarray}
R_b&=\left[R_{\rm conn}^{\left(5-s\right)/2}+
\sqrt{\left(5-s\right)\left(10-3s\right)
\over 8\pi }\left(t-t_{\rm conn}\right)\right]^{2/\left(5-s\right)}\cr
v_b&=\left[{1\over v_b\left(t_{\rm conn}\right)}+{5-s\over 2\left(R_b-R_{\rm conn}\right)
^{\left(s-3\right)/2}}\sqrt{8\pi\over\left(5-s\right)\left(10-3s\right)}\right]
^{-1}\cr
R_r&=\left[{R_{\rm conn}\over l_{\rm ED}t_{\rm conn}}-
v_r\ln\left(t\over t_{\rm conn}\right)\right]t\cr
v_r&={R_{\rm conn}\over l_{\rm ED}t_{\rm conn}}-
{v_b\left(t_{\rm conn}\right)\over l_{\rm ED}},
\end{eqnarray}
where the reverse shock velocity is held constant, similarly to the core
propagation phase in \cite{laming03}. In this approximation, if
$R_r/t < 0.75v_r$, the reverse shocked ejecta flows backwards
towards the remnant center. In this case we put the expansion velocity of reverse
shocked ejecta equal to zero, hence
\begin{equation}
R_r=0.75v_rt=\left[{R_{\rm conn}\over l_{\rm ED}t_{\rm conn}}-
v_r\ln\left(t\over t_{\rm conn}\right)\right]t,
\end{equation}
giving $t=t_{\rm conn}\exp\left(r_{\rm conn}/l_{\rm ED}t_{\rm conn} -0.75\right)$
and so
\begin{equation}
R_r=0.75v_rt_{\rm conn}\exp\left({R_{\rm conn}\over
l_{\rm ED}t_{\rm conn}v_r}-0.75\right),
\end{equation}
and then $v_r=0.75{R_r/t}$. While a little ad hoc, this gives good agreement with
the $s=0$ models given in \citet{truelove99}. We emphasize that the jet ejecta
in which we are most interested undergo reverse shock passage with $t<t_{\rm conn}$,
and so the reverse shock velocity is given accurately by equation A1 and those
in the text immediately following. The later behavior of $v_b$ and $R_b$ is only
necessary for the adiabatic expansion of these shocked knots.

The remaining parameter required is the separation between the forward shock and
the contact discontinuity. In a jet, this can be significantly smaller than the
value obtained in a spherically symmetrical explosion, because plasma
entering the forward shock in the jet region is shocked obliquely, and its
postshock flow is no longer radial. We set this distance equal to zero in our
simulations, so that the contact discontinuity is at the forward shock.

We have also updated the atomic data for dielectronic recombination of K-shell,
and L-shell ions. Recombination from H- to He-like and from He- to Li-like are
taken from \citet{dasgupta04}. The successive isoelectronic sequences Li-,
Be-, B-, C-, N-, O-, and F-like are taken from
\citet{colgan04}\citep[see also][]{colgan05},
\citet{colgan03}, \citet{altun04}\citep[see also][]{altun05},
\citet{zatsarinny04a}, \citet{mitnik04},
\citet{zatsarinny03}\citep[see also][]{zatsarinny05}, and
\citet{gu03} respectively. Additionally dielelectronic
recombination from Ne- to Na-like and from Na- to Mg-like are taken from
\citet{zatsarinny04b} and \citet{gu04}.

\section{The Kompaneets Approximation}
We give a treatment of the Kompaneets approximation for Cas A, following
\citet{bisnovatyi95}, in the case of a uniform explosion into a nonuniform
circumstellar density distribution. We then comment on the modifications
necessary to treat asymmetric explosions into symmetric circumstellar media.
The fundamental assumption leading to the Kompaneets approximation is that for an
explosion into asymmetric media, the pressure behind the blast wave is
constant with position on the shock surface.
Defining the shock surface by $f\left(r,\theta ,t\right)
=0$ in conditions of azimuthal symmetry, then
\begin{equation}
{df\over dt}={\partial f\over\partial t}+{\partial f\over\partial r}{\partial r
\over\partial t}+{\partial f\over\partial\theta}{\partial\theta\over\partial t}
={\partial f\over\partial t}+\nabla f\cdot\vec{v}=0.
\end{equation}
Writing $\left|\nabla f\right|=\sqrt{\left(\partial f/\partial r\right)^2
+\left(\partial f/r\partial\theta\right)^2}$ and $\left|\vec{v}\right|=
\sqrt{\left(\gamma +1\right)P_{sh}/2\rho }=\sqrt{\left(\partial r/\partial t
\right)^2+\left(r\partial \theta/\partial t\right)^2}$ we get,
\begin{equation}
\left(\partial r\over\partial t\right)^2={\left(\gamma +1\right)P_{sh}\over
2\rho}\left[1+\left({1\over r}{\partial r\over\partial\theta}\right)^2\right].
\end{equation}
We assume a density profile $\rho =\rho _0\exp\left(\theta /W\right)\left(
r_0/r\right)^s$, where $\rho _0$ and $r_0$ are the values of $\rho$ and $r$
at some fiducial point, taken here to be the head of the jet, and put
$y=t\sqrt{\left(\gamma +1\right)P_{sh}/2\rho _0}$ so that
\begin{equation}
{r_0^s\over r^s}\left(\partial r\over\partial y\right)^2=
\left[1+\left(\partial\ln r\over
\partial\theta\right)^2\right]\exp\left(-\theta /W\right).
\end{equation}
At this point we remark that a postshock pressure variation of $P=P_0\exp\left(
-\theta /W\right)$ and a uniform ambient density would be mathematically
identical. However such a pressure variation needs to be sustained somehow,
since otherwise with a high postshock sound speed the pressure behind the blast
wave should equilibrate. We have already seen that in the jet region, shocked
circumstellar plasma encounters the blast wave obliquely, and once shocked
flows non-radially away from the jet tip. Consequently one might expect the
dynamics of the forward shock in the jet region to remain ejecta dominated well
past the time that the forward shock elsewhere in the remnant has evolved to
the Sedov-Taylor phase. The cooling time of shocked ejecta at the jet tip is
also sufficiently short to render the sound speed at the jet tip slow, and hence
impede the pressure equilibration among the shocked ejecta. Therefore we will assume
that the expanding jet ``remembers'' the anisotropy of explosion energy
well into the remnant phase and calculate blast wave profiles from the
corresponding modification of the Kompaneets approximation.

We define $\xi = \left(r_0/r\right)^{s/2}\partial r/\partial y
=\partial r_0/\partial y$, evaluated at $\theta =0$ but also assumed
independent of $\theta$ below, rearrange and integrate to find
\begin{equation}
\ln r=\pm\int\sqrt{\xi ^2\exp\left(\theta /W\right)-1}\quad d\theta +\ln\int\xi dy
\end{equation}
and hence
\begin{eqnarray}
{\partial\ln r\over\partial\xi}&=\pm\int{\xi\exp\left(\theta /W\right)\over
\sqrt{\xi ^2\exp\left(\theta /W\right)-1}}d\theta +{\partial\over\partial\xi}
\ln\int\xi dy\cr
&=\pm{2W\over\xi}\left[\sqrt{\xi ^2\exp\left(\theta /W\right)-1}
-\sqrt{\xi ^2-1}\right]+{\partial\over\partial\xi}
\ln\int\xi dy.
\end{eqnarray}
Putting $\xi =\cosh x$ and $\int\xi dy=r_0$ we can integrate to find
\begin{eqnarray}
&{1\over 2W}\ln r/r_0=\cr
&-\sqrt{\xi ^2\exp\left(\theta /W\right)-1}+
\arctan\left(\sqrt{\xi ^2\exp\left(\theta /W\right)-1}\right)
+\sqrt{\xi ^2-1}-
\arctan\left(\sqrt{\xi ^2-1}\right)
\end{eqnarray}
where the -ve sign has been taken. If the normal to the shock front makes an
angle $\alpha$ to the radial direction, then $\partial\ln r/\partial\theta=
\tan\alpha$ and $\tan ^2\alpha =\xi ^2\exp\left(\theta /W\right)-1$.

\citet{lazzati05} give a preferred angular dependence of $dE/d\Omega\propto
\theta ^{-2}$ outside of a central region of constant energy. This is
derived from considerations of the gamma-ray burst jet-cocoon structure
at jet breakout from the stellar envelope, and also reproduces the so-called
Frail correlation. In this case our Kompaneets approximation becomes
\begin{equation}{r_0^s\over r^s}
\left(\partial r\over\partial y\right)^2=\left[1+\left(\partial\ln r\over
\partial\theta\right)^2\right]{W^2\over\theta ^2}.
\end{equation}
Following the steps above we derive
\begin{eqnarray}
\ln {r\over r_0}&=\pm {\xi\over W}\left[{\theta\over 2}
\sqrt{\theta ^2-{W^2\over\xi ^2}} -{W^2\over 2\xi ^2}\ln\left|\theta +
\sqrt{\theta ^2-{W^2\over\xi ^2}}\right|\right]_{\theta =W}^{\theta =\theta}\cr
&=\pm {\xi\over W}\left[{\theta\over 2}
\sqrt{\theta ^2-{W^2\over\xi ^2}} -{W^2\over 2\xi ^2}\ln\left|\theta +
\sqrt{\theta ^2-{W^2\over\xi ^2}}\right|-{W^2\over 2}\sqrt{1-{1\over\xi ^2}}
+{W^2\over 2\xi^2}\ln\left|W+W\sqrt{1-{1\over\xi ^2}}\right|\right].
\end{eqnarray}
Simplified expressions in the limit $1/\xi ^2\rightarrow 0$ or $\xi \sim 1$
and $\theta >>W$ are
$r=r_0\exp\left(-\xi\theta ^2/2W\right)$ and $r=r_0\left(2\theta/W\right)^{W/2}
\exp\left(-\theta ^2/2W\right)$ respectively. \citet{zhang04} prefer $n=3$, and
\citet{graziani05} prefer $n=8$. These cases require numerical integration of the
corresponding version of equation B7.

\clearpage
\begin{deluxetable}{cccccccccc}
\tabletypesize{\footnotesize}
\tablecaption{NEI Models for Jet Tip Knots with O Continuum}
\tablewidth{0pt}
\tablehead{
\colhead{Region}&\colhead{Counts}&\colhead{Region Size}&\colhead{$\chi^2/dof$}&\colhead{N$_{\rm H}$}&\colhead{kT}&\colhead{n$_{\rm e}$t}&\colhead{Si}&\colhead{Fe}\\
}
\startdata
N tip & 3134 & $4.2\times 2.2$ & 78.5, 1.11 & 1.2 & 0.60 & 1.6e13 & 2200 & 640 \\
 & & & & & (0.55-0.66) & ($>$1.4e13) & (1170-4000) & (250-1600) \\
M tip & 9626 & $9.5\times 2.7$ & 187.9, 1.36 & 1.3 & 0.73 & 9.6e12 & 2.3 & 0.44 \\
 & & & & & (0.70-0.75) & ($>$4.6e12) & (1.6-2.6) & (0.26-0.59) \\
S tip & 2895 & r=1.5 & 92.2, 1.40 & 1.38 & 0.60 & 9.4e12 & 3.1 & 4.2 \\
 & & & & (1.30-1.46) & (0.58-0.62) & ($>$ 6.5e12) & (1.5-672) & (2.1-673) \\
\enddata
\end{deluxetable}

\begin{deluxetable}{cccccccc}
\tabletypesize{\footnotesize}
\tablecaption{Line Strengths in Jet Knots}
\tablewidth{0pt}
\tablehead{
\colhead{Region}&\colhead{Counts}&\colhead{Region Size}&\colhead{Fe K centroid}&\colhead{Fe K EQW}&\colhead{Ca He $\alpha$ EQW }&\colhead{Si He/Ly ratio}&\colhead{kT}\\
}
\startdata
\cutinhead{North Filament}
jet3 & 19953 & $5.7\times 4.9$ & 6.632 & 4.5 (3.6-5.3) & 0.99 (0.83-1.2) & 2.7 (2.0-3.5) & 0.74,3.4 \\
j4 & 38989 & $9.6\times 2.1$ & 6.635 & 3.7 (3.1-4.3) & 0.76 (0.71-0.83) & 3.3 (3.1-3.6) & [0.17], 2.4 \\
j5e & 19120 & $4.4\times 4.2$ & 6.690 & 2.4 (1.3-3.3) & 1.0 (0.81-1.3) & 4.2 (3.2-4.8) & 0.45, 3.7 \\
j5a & 19279 & $7.5\times 3.5$ & 6.643 & 2.6 (2.1-3.2) & 0.88 (0.68-0.82) & 2.2 (1.9-2.4) & 1.0, 11\\
j5c & 32972 & $7.2\times 3.2$ & 6.722 & 2.5 (2.4-2.6) & 0.53 (0.45-0.62) & 5.7 (4.9-6.7) & 0.41, 2.1 \\
j5b & 27203 & $8.4\times 2.1$ & [6.661, 0.05] & 0.9 (0.07-1.7) & 0.19 (0.13-0.30) & 3.9 (3.0-4.7) & 0.58, 1.8 \\
\cutinhead{Middle Filament}
j8 & 23291 & $7.6\times 4.0$ & 6.592 & 2.8 (2.0-3.6) & 1.2 (1.1-1.4) & 6.3 (5.5-7.3) & 0.46, 3.7 \\
j10 &37045 & $9.1\times 2.4$ & 6.601 &2.3 (1.5-3.1) & 1.0 (0.91-1.2) & 6.0 (5.6-6.7) & 0.55,2.6 \\
j11 & 25017 & $7.0\times 2.6$ & 6.666 & 2.0 (1.4-2.7) & 1.1 (0.93-1.2) & 3.5 (3.0-4.0) & 0.53, 3.7 \\
j11a & 29189 & $9.2\times 3.8 $ & 6.626 & 2.1 (1.6-2.7) & 0.64 (0.55-0.72) & 3.3 (3.0-3.8) & 0.61, 2.9 \\
\cutinhead{South Filament}  
j13a & 39392 & $4.6\times 2.2$ & 6.605 & 1.7 (1.1-2.1) & 0.76 (0.68-0.83) & 4.2 (4.1-4.5) & 1.8, 80 \\
j13b & 20171 & $5.3\times 1.6$ & [6.610, 0.1] & 1.3 (0.3-1.7) & 0.43 (0.36-0.51) & 6.6 (6.0-7.7) &  \\
j18 & 74142 & $11.2\times 2.9$ & 6.645 & 0.58 (0.44-0.73) & 0.10 (0.08-0.12) & 4.0 (3.8-4.3) & 1.7, [150]\\
j19 & 62144 & $8.0\times 2.8$ & 6.686 & 1.2 (1.0-1.4) & 0.17 (0.15-0.20) & 3.1 (3.0-3.3) & 1.7, [150] \\
j20 & 44787 & $6.0\times 2.3$ & 6.676 & 1.7 (1.3-2.1) & 0.22 (0.18-0.28) & 2.8 (2.7-3.0) & 1.6, [100] \\
j21 & 60072 & $9.7\times 2.6$ & 6.648 & 1.3 (0.95-1.5) & 0.11 (0.09-0.14) & 2.7 (2.5-2.8) & 1.6, [150] \\
\cutinhead{Counter-Jet}
cjet & 60518 & & 6.597 & 4.0 (3.0-5.1) & 0.37 (0.30-0.45) & 4.4 (3.4-5.3) & 0.39, 1.86 \\
\enddata
\end{deluxetable}

\begin{deluxetable}{cccccccc}
\tabletypesize{\small}
\tablecaption{{\em pshock} Spectral Fits for Jet Knots with O continuum}
\tablewidth{0pt}
\tablehead{
\colhead{Region}&\colhead{gain factor}&\colhead{$\chi^2/dof$}&\colhead{N$_{\rm H}$}&\colhead{kT}&\colhead{n$_{\rm e}$t}&\colhead{Si}&\colhead{Fe}\\
}
\startdata
\cutinhead{North Filament---pshock}
jet3 & 1.001 & 254.4, 1.30 & 1.57 & 2.32 & 2.1e11 & 1.2 & 0.68 \\
 & & & (1.51-1.64) & (2.09-2.56) & (1.8-2.6e11) & (0.96-1.5) & (0.56-0.90) \\
j4 & 0.999 & 338.4, 1.54 & 1.29 & 2.00 & 2.4e11 & 1.1 & 0.58 \\
 & & & (1.28-1.32) & (1.89-2.13) & (2.2-2.7e11) & (1.0-1.2) & (0.52-0.67) \\
j5b & 0.997 & 224.3, 1.32 & 1.18 & 1.26 & 5.6e11 & 0.56 & 0.23 \\
 & & & (1.13-1.21) & (1.15-1.48) & (3.4-9.0e11) & (0.48-0.61) & (0.19-0.26) \\
j5e & 0.999 & 194.8, 1.11 & 1.20 & 1.74 & 2.5e11 & 0.96 & 0.52 \\
 & & & (1.14-1.21) & (1.59-1.94) & (2.1-2.9e11) & (0.84-1.0) & (0.40-0.55) \\
j5c & 0.999 & 428.2, 2.25 & 1.18 & 1.40 & 2.4e11 & 1.1 & 0.50 \\
j5a & 0.999 & 347.7, 1.61 & 1.17 & 2.79 & 2.1e11 & 0.68 & 0.28 \\
 & & & (1.14-1.20) & (2.51-2.96) & (1.9-2.5e11) & (0.58-0.72) & (0.24-0.30) \\
\cutinhead{North Filament---NEI}
jet3 & 1.001 & 350.6, 1.79 & 1.65 & 2.15 & 1.1e11 & 2.1 & 1.0 \\
 & & & (1.59-1.70) & (1.93-2.35) & (9.4e10-1.2e11) & (1.2-18) & (0.75-38) \\
j4 & 1.000 & 670.8, 3.05 & 1.25 & 1.85 & 1.1e11 & 1.1 & 0.51 \\
j5b & 0.997 & 273.2, 1.61 & 0.86 & 1.07 & 5.1e11 & 0.45 & 0.91 \\
 & & & (0.77-0.96) & (1.03-1.15) & (3.5-6.4e11) & (0.38-0.49) & (0.07-0.12) \\
j5e & 0.999 & 275.9, 1.57 & 0.81 & 1.15 & 3.8e11 & 0.54 & 0.13 \\
 & & & (0.75-0.85) & (1.13-1.21) & 2.7-4.5e11) & (0.46-0.58) & (0.10-0.14) \\
j5c & 1.001 & 693.4, 3.65 & 1.03 & 1.24 & 1.2e11 & 0.73 & 0.24 \\
j5a & 1.003 & 589.0, 2.73 & 1.04 & 3.55 & 6.0e10 & 0.49 & 0.18 \\
\cutinhead{Middle Filament---pshock}
j8 & 0.998 & 240.0, 1.26 & 1.49 & 1.96 & 1.5e11 & 1.5 & 0.58 \\
 & & & (1.45-1.54) & (1.91-2.17) & (1.4-1.6e11) & (1.3-1.9) & (0.50-0.70) \\
j10 & 0.999 & 285.7, 1.43 & 1.28 & 1.62 & 1.8e11 & 1.8 & 0.69 \\
 & & & (1.26-1.30) & (1.53-1.79) & (1.6-2.3e11) & (1.5-1.9) & (0.52-0.71) \\
j11 & 0.998 & 254.8, 1.32 & 1.22 & 1.83 & 2.7e11 & 1.5 & 0.50 \\
 & & & (1.16-1.26) & (1.68-2.03) & (2.3-3.4e11) & (1.2-1.8) & (0.41-0.61) \\
j11a & 1.000 & 270.3, 1.26 & 1.26 & 1.86 & 2.5e11 & 0.73 & 0.26 \\
 & & & (1.21-1.28) & (1.76-1.98) & (2.2-3.1e11) & (0.66-0.76) & (0.20-0.28) \\
\cutinhead{Middle Filament---NEI}
j8 & 0.998 & 315.3, 1.66 & 1.33 & 1.85 & 7.8e10 & 1.1 & 0.33 \\
 & & & (1.28-1.39) & (1.58-2.00) & (7.0e10-1.0e11) & (0.90-1.2) & (0.22-0.36) \\
j10 & 0.998 & 421.4, 2.11 & 1.20 & 0.99 & 4.2e11 & 2.0 & 0.57 \\
j11 & 0.998 & 348.4, 1.81 & 0.86 & 1.20 & 3.9e11 & 0.89 & 0.15 \\
 & & & (0.80-0.92) & (1.16-1.26) & (3.2-4.6e11) & (0.78-1.0) & (0.13-0.20) \\
j11 & 0.999 & 375.3, 1.75 & 0.79 & 1.26 & 3.4e11 & 0.51 & 0.066 \\
 & & & (0.78-0.81) & (1.24-1.37) & (3.2-3.9e11) & (0.49-0.52) & (0.057-0.074) \\
\cutinhead{South Filament---pshock}
j13a & 1.006 & 358.4, 1.66 & 1.39 & 2.12 & 1.5e11 & 1.7 & 0.26 \\
  & & & (1.35-1.43) & (2.05-2.30) & (1.3-1.6e11) & (1.6-1.9) & (0.23-0.31) \\
j13b & 1.009 & 229.6, 1.26 & 1.50 & 1.96 & 1.0e11 & 1.2 & 0.24 \\
  & & & (1.47-1.55) & (1.79-2.13) & (9.2e10-1.1e11) & (1.1-1.3) & (0.20-0.26) \\
j18 & 1.008 & 493.6, 1.61 & 1.47 & 1.95 & 1.8e11 & 0.28 & 0.11 \\
 & & & (1.46-1.49) & (1.85-1.98) & (1.7-1.9e11) & (0.27-0.29) & (0.10-10.) \\
j19 & 1.008 & 518.9, 1.87 & 1.37 & 2.32 & 1.8e11 & 0.50 & 0.19 \\
 & & & (1.35-1.39) & (2.22-2.42) & (1.7-1.9e11) & (0.47-0.52) & (0.17-0.20) \\
j20 & 1.008 & 450.8, 1.91 & 1.21 & 2.27 & 2.1e11 & 0.74 & 0.18 \\
  & & & (1.19-1.25) & (2.11-2.40) & (1.9-2.4e11) & (0.68-0.78) & (0.15-0.20) \\
j21 & 1.008 & 421.9, 1.60 & 1.20 & 1.68 & 4.3e11 & 0.59 & 0.15 \\
 & & & (1.17-1.22) & (1.61-1.76) & (3.9-4.9e11) & (0.57-0.63) & (0.13-0.16) \\
\cutinhead{South Filament---NEI}
j13a & 1.005 & 469.5, 2.16 & 1.21 & 2.08 & 7.3e10 & 1.1 & 0.15 \\
j13b & 1.004 & 308.3, 1.69 & 1.36 & 1.85 & 5.8e10 & 0.88 & 0.14 \\
 & & & (1.31-1.40) & (1.73-2.04) & (5.3-6.7e10) & (0.79-0.92) & (0.11-0.15) \\
j18 & 1.007 & 731.9, 2.39 & 1.27 & 2.02 & 7.3e10 & 0.23 & 0.075 \\
j19 & 1.008 & 807.6, 2.92 & 1.20 & 2.64 & 6.6e10 & 0.39 & 0.13 \\
j20 & 1.008 & 771.9, 3.27 & 1.02 & 2.87 & 6.6e10 & 0.56 & 0.10 \\
j21 & 1.008 & 567.7, 2.16 & 0.72 & 1.58 & 2.0e11 & 0.37 & 0.037 \\
\cutinhead{Counter-Jet---pshock}
cjet & 1.000 & 315.7, 1.36 & 2.06 & 1.26 & 4.1e11 & 1.22 & 1.2 \\
 & & & (2.04-2.11) & (1.24-1.29) & (3.8-4.5e11) & (1.17-1.43) & (0.97-1.43) \\
\cutinhead{Counter-Jet---NEI}
cjet & 1.000 & 427.6, 1.57 & 1.95 & 1.08 & 2.9e11 & 0.79 & 0.63 \\
 & & & (1.93-2.03) & (1.05-1.09) & (2.6-3.9e11) & (0.74-1.04) & (0.56-1.00) \\
\enddata
\end{deluxetable}

\clearpage

\begin{figure}
\label{regions} \epsscale{1.0} \plotone{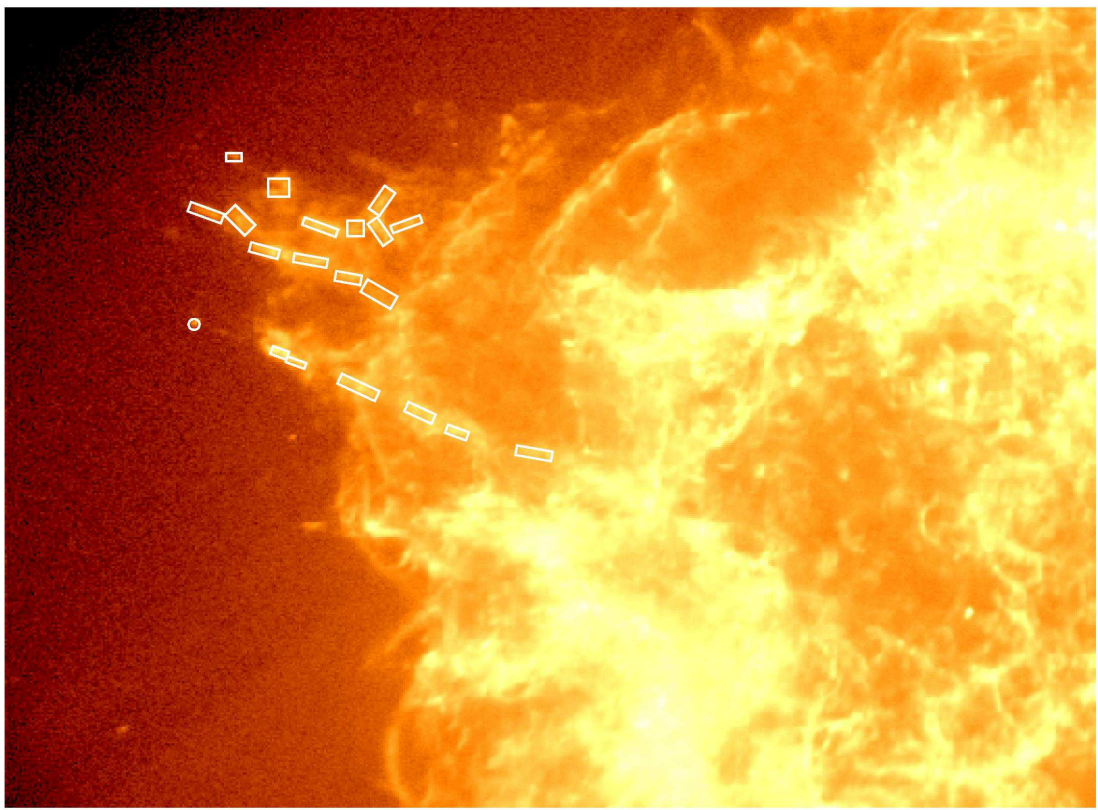} \figcaption{Regions
of the three filaments (north, middle, south) of the northeast jet
used for spectral analysis.}
\end{figure}

\clearpage

\begin{figure}
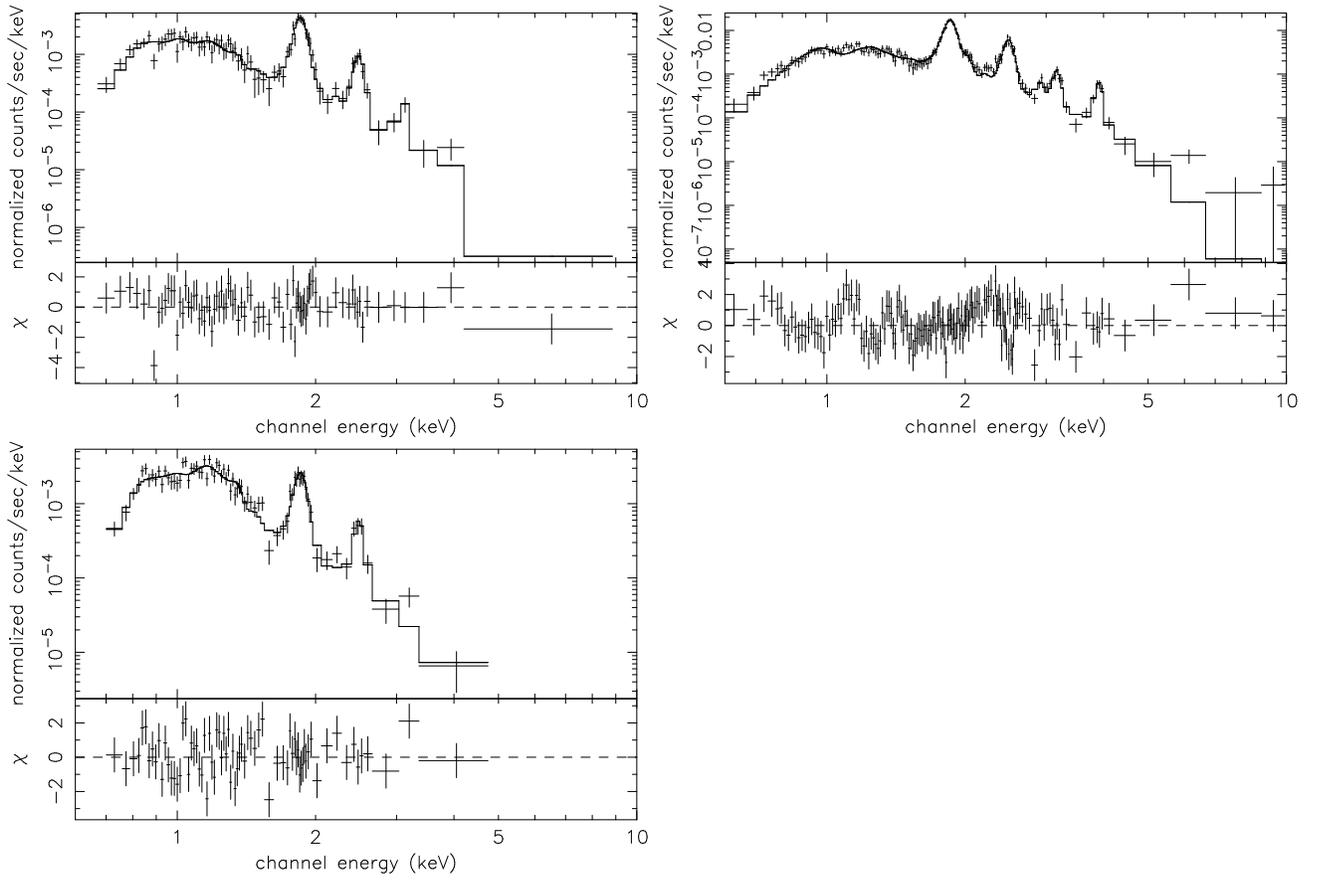

\label{spectra}
\includegraphics[scale=0.35,angle=-90]{f2a.ps}
\includegraphics[scale=0.35,angle=-90]{f2b.ps}
\includegraphics[scale=0.35,angle=-90]{f2c.ps}
\figcaption{Spectra of the outermost jet knots (north at upper left,
middle at upper right, south at bottom).}
\end{figure}

\begin{figure}
\label{eqw}
\includegraphics[scale=0.4]{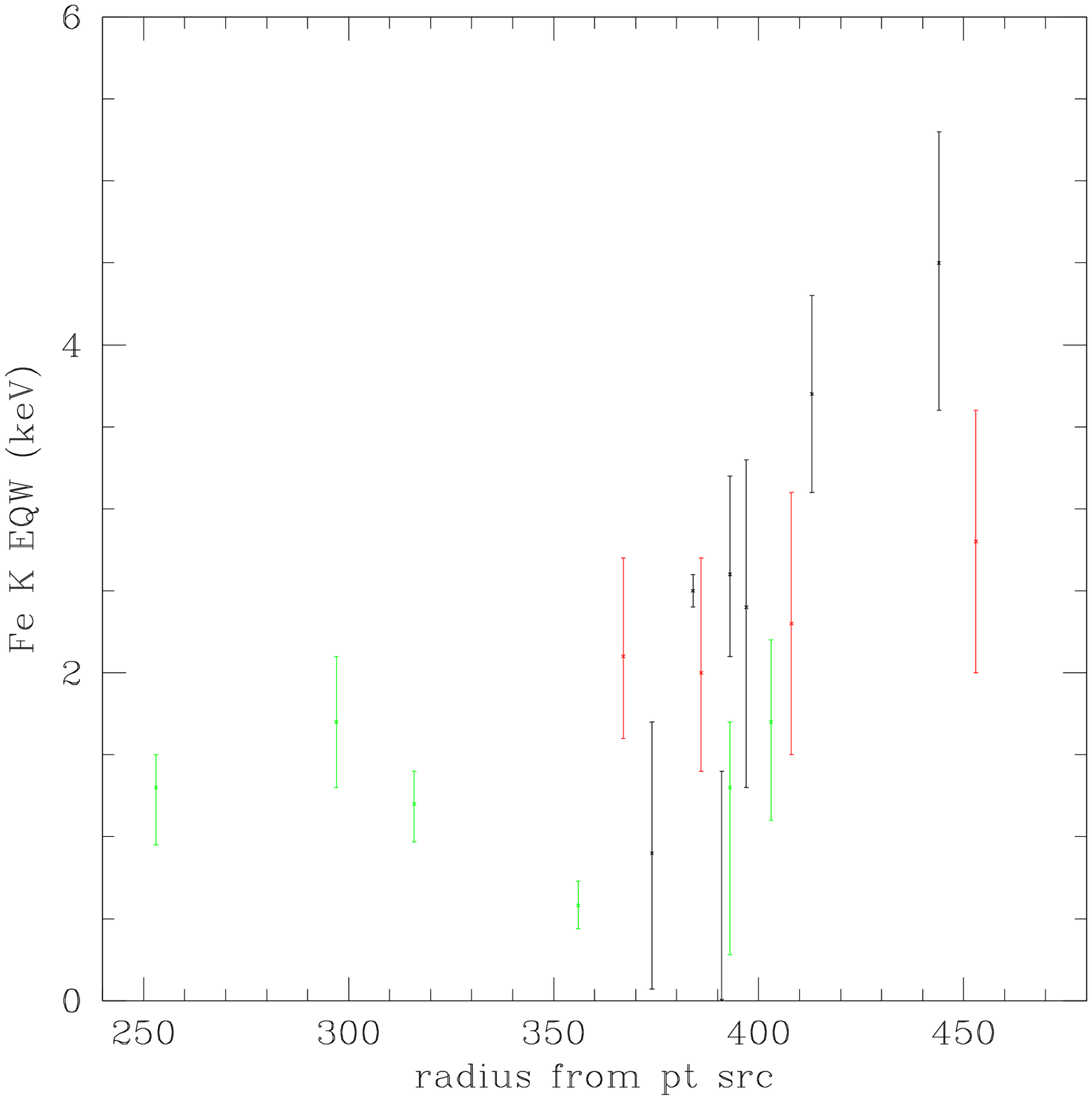}\includegraphics[scale=0.4]{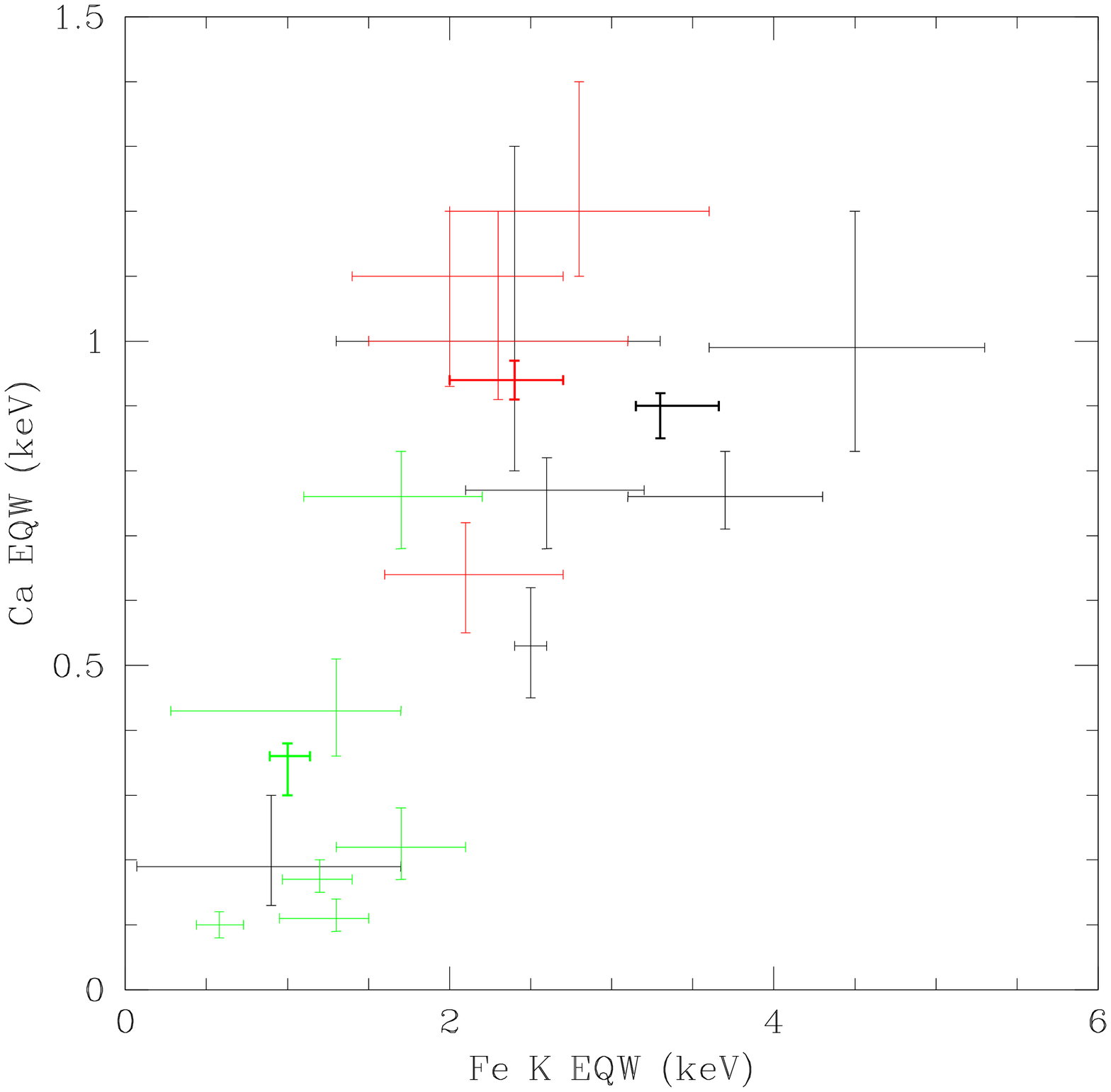}
\figcaption{The Fe K equivalent width plotted for knots in the
north (black), middle (red), and south (green) jet filaments as a
function of the distance of the center of the extraction region
relative to the position of the Cas A point source near the center of
the remnant.  The correlation of Ca EQW compared to Fe EQW is shown in
the panel to the right.}
\end{figure}

\begin{figure}
\label{cavities}
\epsscale{0.75}
\plotone{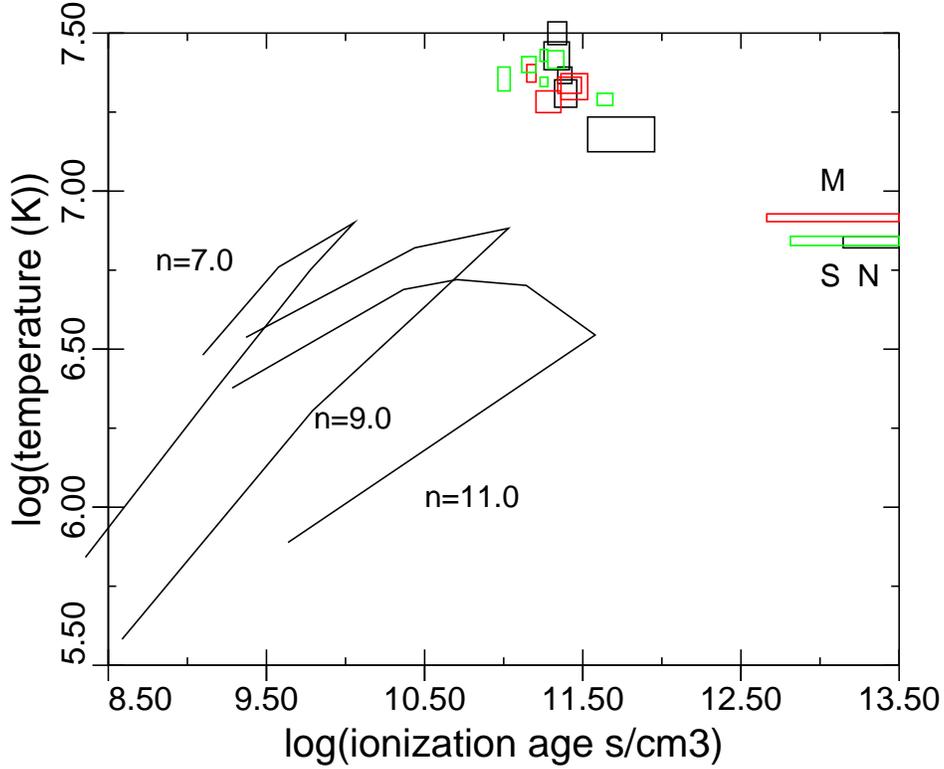}
\figcaption[f4.eps]{The loci on the electron temperature - ionization
age plane of fitted knots in the NE jet, color coded such that knots from the
middle filament are red, knots from the south filament are green, and those
from the north filament are black. Also plotted are curves derived from models
or circumstellar cavities for ejecta profiles corresponding to a uniform
density core and a power law outer envelope, $\rho\propto r^{-n}$, with
$n=7$, 9, and 11. The discrepancy between observations and models suggests that
CSM cavity models cannot be responsible for the jet morphology.}
\end{figure}

\begin{figure}
\label{jets}
\epsscale{0.75}
\plotone{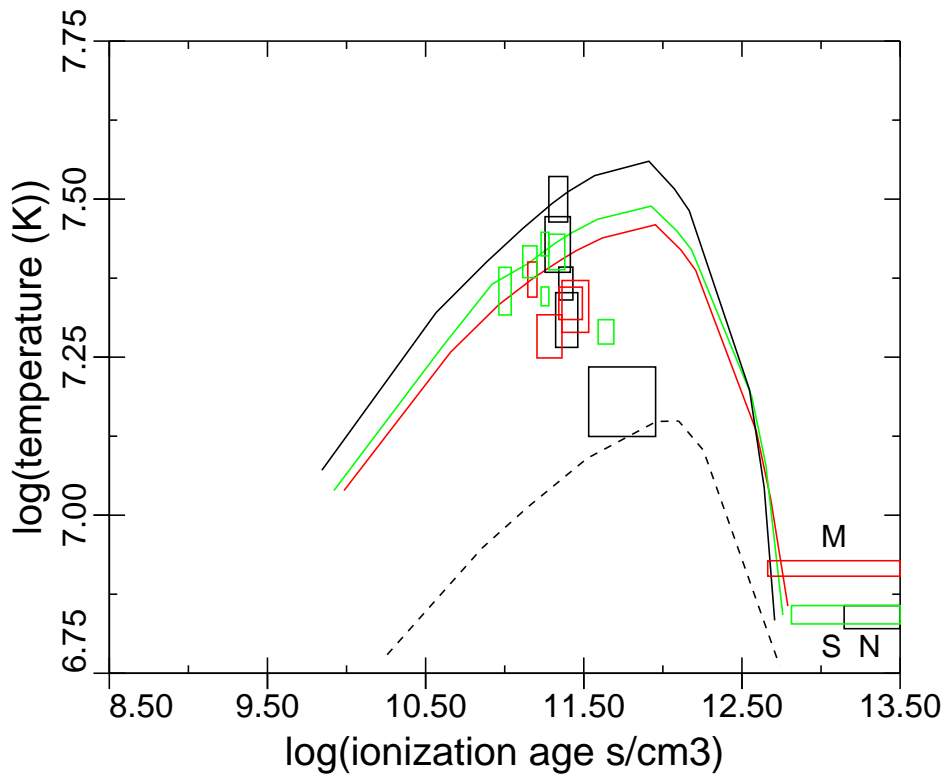}
\figcaption[f5.eps]{The same fitted jet knots as in Figure 4, compared with
jet models specified to match the fitted abundances in the knots at the jet
tips, as solid lines with the same color coding. Small differences in the onset
of thermal instability are visible, due to the different abundance sets. The
black dashed line shows the result of assuming half the plasma in the N filament
to be H dominated. This reduces the electron temperature by over a factor of
two, but the onset of thermal instability is less sudden. Jet models appear to
match the data points much better.}
\end{figure}

\begin{figure}
\label{fe_si}
\epsscale{0.6}
\plotone{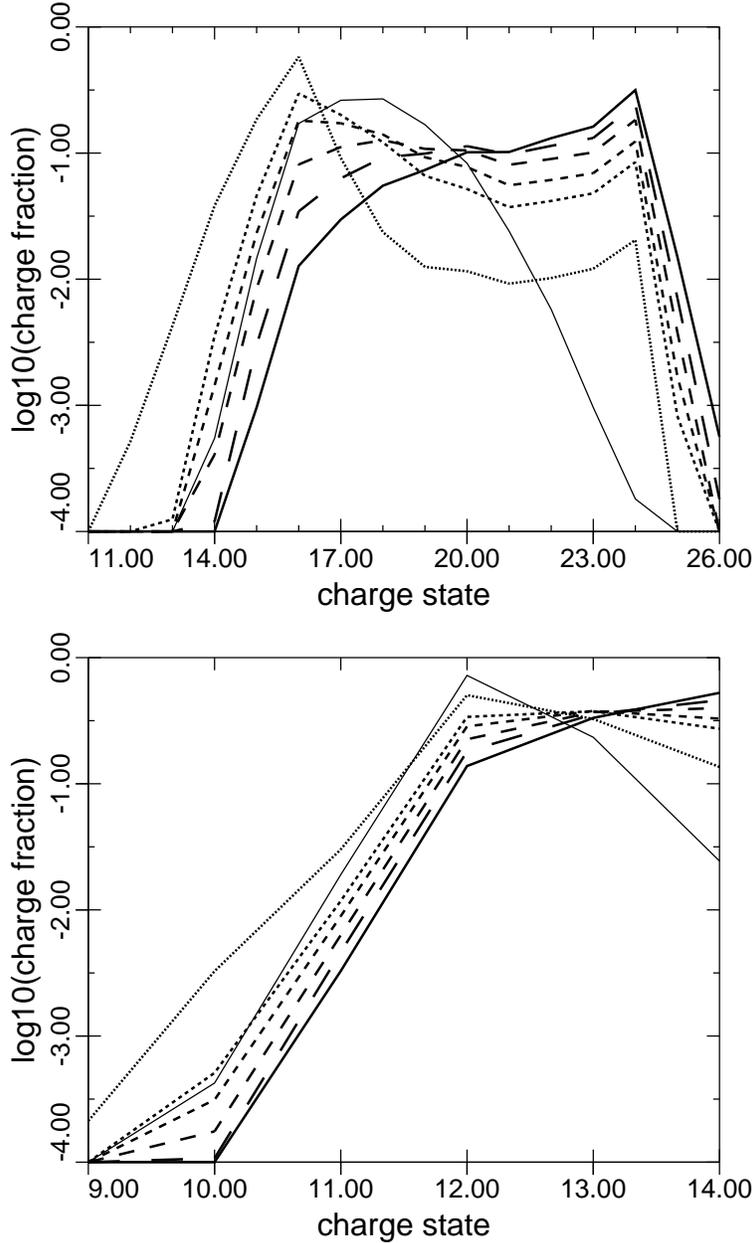}
\figcaption[f6.eps]{Charge state distributions of Fe (top) and Si (bottom)
corresponding to ejecta encountering the reverse shock at different times
following explosion. The thick solid line indicates Fe (or Si) going through
the reverse shock 1.6 year after explosion, which gives the best agreement
with the fitted electron temperature. The dashed curves with sucessively
finer and finer dashes are
for ejecta undergoing reverse shock passage 1.45, 1.4, 1.35, 1.315, and 1.3
years after explosion, with electron temperatures $7.84\times 10^6$,
$7.04\times 10^6$, $6.17\times 10^6$,
$5.51\times 10^6$ and $5.22\times 10^6$K respectively.
Although these ejecta cool to lower temperatures than
indicated by the fits, the charge state distributions are in better agreement
with the observations. The narrow solid lines indicate the collisional
ionization equilibrium charge state distributions for a temperature of
$8.6\times 10^6$ K, and active indicative of the actual ionization balance
detected in the jet knots.}
\end{figure}

\begin{figure}
\label{profiles}
\epsscale{1.00}
\plotone{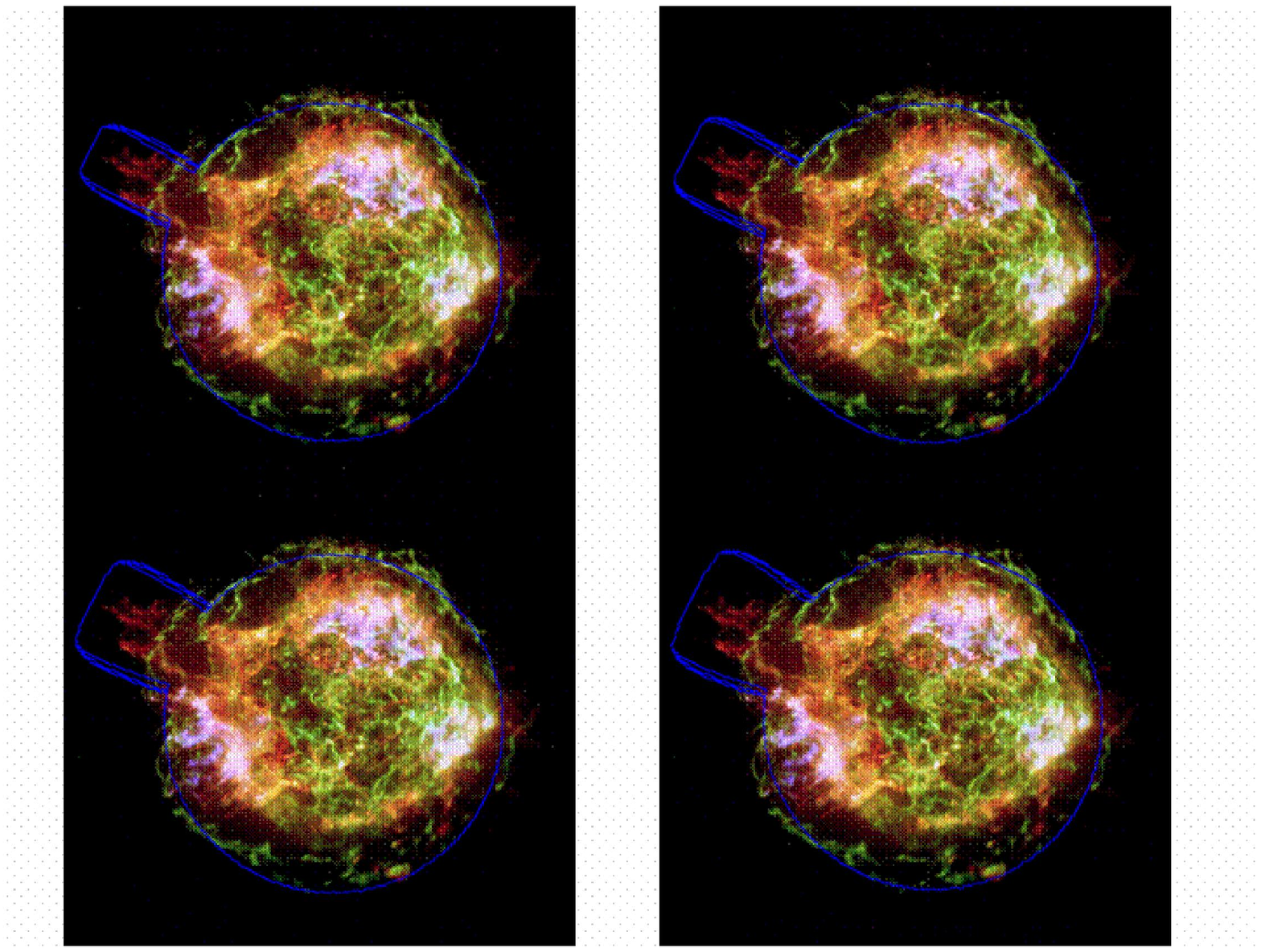}
\figcaption[f7.ps]{Illustrative plots of the blast wave profile in the
NE jet region for jet models with a uniform energy core and power law
outer region, $E\propto \theta ^{-n}$, with $n=2$, 3, and 8,
\citep[based on][respectively]{lazzati05,zhang04,graziani05} for uniform
energy core opening angles of 5 degrees (top left), 7 degrees (top right),
9 degrees (bottom left) and 11 degrees (bottom right).}
\end{figure}


\begin{thebibliography}{}
\bibitem[Akiyama et al.(2003)]{akiyama03}Akiyama, S., Wheeler, J. C., Meier,
D. L., \& Lichtenstadt, I. 2003, \apj, 584, 954
\bibitem[Altun et al.(2004)]{altun04}Altun, Z., Yumak, A., Badnell, N. R.,
Colgan, J., \& Pindzola, M. S. 2004, \aap, 420, 775
\bibitem[Altun et al.(2005)]{altun05}Altun, Z., Yumak, A., Badnell, N. R.,
Colgan, J., \& Pindzola, M. S. 2005, \aap, 433, 395
\bibitem[Atoyan, Buckley \& Krawczynski(2005)]{atoyan05}Atoyan, A., Buckley, J.,
\& Krawczynski, H. 2005, astro-ph/05096115
\bibitem[Bell \& Lucek(2001)]{bell01}Bell, A. R., \& Lucek, S. G. 2001,
\mnras, 321, 433
\bibitem[Bisnovatyi-Kogan \& Silich (1995)]{bisnovatyi95} Bisnovatyi-Kogan, G. S.,
\& Silich, S. A. 1995, Rev. Mod. Phys. 67, 661
\bibitem[Blondin, Lundqvist, \& Chevalier(1996)]{blondin96}Blondin, J. M.,
Lundqvist, P., \& Chevalier, R. A. 1996, \apj, 472, 257
\bibitem[Blue et al.(2005)]{blue05}Blue, B. E. 2005, \prl, 94, 095005
\bibitem[Butler et al.(2005)]{butler05}Butler, N., Ricker, G., Vanderspek, R.,
Ford, P., Crew, G., Lamb, D. Q., \& Jernigan, J. G. 2005, \apj, 627, L9
\bibitem[Chevalier \& Li(2000)]{chevalier00}Chevalier, R. A., \& Li, Z. 2000,
\apj, 536, L57
\bibitem[Chevalier \& Oishi(2003)]{chevalier03}Chevalier, R. A., \& Oishi, J. 2003,
\apj, 593, L23
\bibitem[Colgan et al.(2003)]{colgan03}Colgan, J., Pindzola, M. S., Whiteford,
A. D., \& Badnell, N. R. 2003, \aap, 412, 597
\bibitem[Colgan, Pindzola, \& Badnell(2004)]{colgan04}Colgan, J., Pindzola,
M. S., \& Badnell, N. R. 2004, \aap, 417, 1183
\bibitem[Colgan, Pindzola, \& Badnell(2005)]{colgan05}Colgan, J., Pindzola,
M. S., \& Badnell, N. R. 2005, \aap, 429, 369
\bibitem[Dasgupta \& Whitney(2004)]{dasgupta04}Dasgutpa, A., \& Whitney, K. G.
2004, PRA, 69, 022702
\bibitem[Delaney et al.(2004)]{delaney04}Delaney, T., Rudnick, L., Fesen, R. A.,
Jones, T. W., Petre, R., \& Morse, J. 2004, \apj, 613, 343
\bibitem[Drury(1983)]{drury83}Drury, L. O'C. 1983, Rep. Prog. Phys. 46, 973
\bibitem[Fesen(2001)]{fesen01}Fesen, R. A. 2001, \apjs, 133, 161
\bibitem[Folatelli et al.(2005)]{folatelli05}Folatelli, G., et al. 2005,
astro-ph/0509731
\bibitem[Foster et al.(2005)]{foster05}Foster, J. M., et al. 2005, \apj, 634, L77
\bibitem[Frail et al.(2001)]{frail01}Frail, D. A., et al. 2001, \apj, 562, L55
\bibitem[Galama et al.(1998)]{galama98}Galama, T. J., et al. 1998, Nature, 395,
670
\bibitem[Garnavich et al.(2003)]{garnavich03}Garnavich, P. M., et al. 2003, \apj,
582, 924
\bibitem[Graziani, Lamb, \& Donaghy(2005)]{graziani05}Graziani, C., Lamb, D. Q.,
\& Donaghy, T. Q. 2005, \apj, submitted, astro-ph/0505623
\bibitem[Gu(2003)]{gu03}Gu, M. F. 2003, \apj, 590, 1131
\bibitem[Gu(2004)]{gu04}Gu, M. F. 2004, \apjs, 153, 389
\bibitem[Hjorth et al.(2003)]{hjorth03}Hjorth, J., et al. 2003, Nature, 423, 847
\bibitem[Hwang \& Laming(2003)]{hwang03}Hwang, U., \& Laming, J. M. 2003, \apj,
597, 362
\bibitem[Hwang et al.(2000)]{hwang00}Hwang, U., Holt, S. S., \& Petre, R.
2000, \apj, 537, L119
\bibitem[Hwang et al.(2004)]{hwang04}Hwang, U. et al. 2004, \apj, 615, L117
\bibitem[Iyudin et al.(1994)]{iyudin94}Iyudin, A. F., et al. 1994,
\aap, 284, L1
\bibitem[Jun, Jones \& Norman(1996)]{jun96}Jun, B.-I., Jones, T. W., \& Norman,
M. L. 1996, \apj, 468, L59
\bibitem[Khokhlov et al.(1999)]{khokhlov99}Khokhlov, A. M., H\"oflich, P. A.,
Oran, E. S., Wheeler, J. C., Wang, L., \& Chtchelkanova, A. Yu. 1999, \apj, 524,
L107
\bibitem[Kifonidis et al.(2000)]{kifonidis00}Kifonidis, K., Plewa, T., Janka, H.-T.,
\& M\"uller, E. 2000, \apj, 531, L123
\bibitem[Kifonidis et al.(2003)]{kifonidis03}Kifonidis, K., Plewa, T., Janka, H.-T.,
\& M\"uller, E. 2003, \aap, 408, 621
\bibitem[Laming \& Hwang(2003)]{laming03}Laming, J. M., \& Hwang, U. 2003, \apj,
597, 347
\bibitem[Laming \& Grun(2002)]{laming02}Laming, J. M., \& Grun, J. 2002, Phys.
Rev. Lett., 89, 125002
\bibitem[Lazzati \& Begelman(2005)]{lazzati05}Lazzati, D., \& Begelman, M. C.
2005, \apj, 629, 903
\bibitem[Lebedev et al.(2005)]{lebedev05}Lebedev, S. V. et al. 2005, \mnras,
361, 97
\bibitem[Lucek \& Bell(2001)]{lucek01}Lucek, S. G., \& Bell, A. R. 2001,
\mnras, 314, 65
\bibitem[Morse et al.(2004)]{morse04}Morse, J. A., Fesen, R. A., Chevalier, R. A.,
Borkowski, K. J., Gerady, C. L., Lawrence, S. S., \& van den Bergh, S. 2004,
\apj, 614, 727
\bibitem[MacFayden \& Woosley(1999)]{macfayden99}MacFayden, A. I., \& Woosley,
S. E. 1999, \apj, 524, 262
\bibitem[Malesani et al.(2004)]{malesani04}Malesani, D., et al. 2004, \apj, 609,
L5
\bibitem[Mazzali et al.(2002)]{mazzali02}Mazzali, P., et al. 2002, \apj,572, L61
\bibitem[Mazzali et al.(2005)]{mazzali05}Mazzali, P., et al. 2005, Science,
308, 1284
\bibitem[Mitnik \& Badnell(2004)]{mitnik04}Mitnik, D. M., \& Badnell, N. R.
2004, \aap, 425, 1153
\bibitem[Nagataki et al.(1998)]{nagataki98}Nagataki, S., Hashimoto, M., Sato, K.,
Yamada,S., \& Mochizuki, Y. 1998, \apj, 492, L45
\bibitem[Nagataki et al.(2003)]{nagataki03}Nagataki, S., Mizuta, A., Yamada, S.,
Takabe, H., \& Sato, K. 2003, \apj, 596,401
\bibitem[Nagataki, Mizuta \& Sato(2006)]{nagataki06}Nagataki, S., Mizuta, A.,
\& Sato, K. 2006, \apj, submitted, astro-ph/0601111
\bibitem[Paczynski(1998)]{paczynski98}Paczynski, B. 1998, \apj, 494, L45
\bibitem[Panaitescu \& Kumar(2002)]{panaitescu02}Panaitescu, A., \& Kumar, P.
2002, \apj, 571, 779
\bibitem[Piro et al.(2005)]{piro05}Piro, L., et al. 2005, \apj, 623, 314
\bibitem[Podsiadlowski et al.(2004)]{podsiadlowski04}Podsiadlowski, Ph.,
Mazzali, P. A., Nomoto, K., Lazzati, D., \& Cappellaro, E. 2004, \apj, 607, L17
\bibitem[Proga et al.(2003)]{proga03}Proga, D., MacFadyen, A. I., Armitage, P. J.,
\& Begelman, M. C. 2003, \apj, 577, L5
\bibitem[Proga(2005)]{proga05}Proga, D. 2005, \apj, 629, 397
\bibitem[Ramirez-Ruiz \& Madau(2004)]{ramirez04}Ramirez-Ruiz, E., \& Madau, P.
2004, \apj, 608, L89
\bibitem[Reeves et al.(2002)]{reeves02}Reeves, J. N. et al. 2002, Nature,
416, 512
\bibitem[Reeves et al.(2003)]{reeves03}Reeves, J. N., Watson, D., Osbourne, J.
P., Pounds, K. A., \& O'Brien, P. T. 2003, \aap, 403, 463
\bibitem[Rutledge \& Sako(2003)]{rutledge03}Rutledge, R. E., \& Sako, M. 2003,
MNRAS, 339, 600
\bibitem[Sako, Harrison \& Rutledge(2005)]{sako05}Sako, M., Harrison, F. A., \&
Rutledge, R. E. 2005, \apj, 623, 973
\bibitem[Sch\"onfelder et al.(2000)]{schonfelder00}Schonfelder,
V.,et al. 2000, in AIP Conf. Proc. 510, 5th Compton Symp., ed. M. L.
McConnell \& J. M. Ryan (Melville: AIP), 54
\bibitem[Soderberg et al.(2006)]{soderberg06}Soderberg, A. M., et al. 2006, \apj,
636, 391
\bibitem[Sonneborn et al.(1998)]{sonneborn98}Sonneborn, G., et al. 1998, \apj,
492, L139
\bibitem[Thorstensen, Fesen, \& van den Bergh(2001)]{thorstensen01}Thorstensen,
J. R., Fesen, R. A., \& van den Bergh, S. 2001, \aj, 122, 297
\bibitem[Tominaga et al.(2005)]{tominaga05}Tominaga, N., et al. 2005, \apj, 633, L97
\bibitem[Truelove \& McKee(1999)]{truelove99}Truelove, J. K., \& McKee, C. F.
1999, \apjs, 120, 299
\bibitem[van Putten(2004)]{vanputten04}van Putten, M. H. P. M. 2004, \apj, 611,
L81
\bibitem[Vink et al.(2001)]{vink01}Vink, J., Laming, J. M., Kaastra, J. S.,
Bleeker, J. A. M., Bloemen, H., \& Oberlack, U. 2001, \apj, 560, L79
\bibitem[Vink \& Laming(2003)]{vink03}Vink, J., \& Laming, J. M. 2003, \apj,
584, 758
\bibitem[Wang et al.(2002)]{wang02}Wang, L., et al. 2002, \apj, 579, 671
\bibitem[Wang(1999)]{wang99}Wang, Q. D. 1999, \apj, 517, L27
\bibitem[Wang, Lai, \& Han(2005)]{wang05}Wang, C., Lai, D., \& Han, J. L. 2005,
astro-ph/0509484
\bibitem[Watson et al.(2002)]{watson02}Watson, D., Jeeves, J. N., Osbourne, J. S.,
O'Brien, P. T., Pounds, K. A., Tedds, J. A., Santos-Lle\'o, M., \& Ehle, M.
2002, \aap, 393, L1
\bibitem[Watson et al.(2003)]{watson03}Watson, D., Reeves, J. N., Hjorth, J.,
Jakobsson, P., \& Pedersen, K. 2003, \apj, 595, L29
\bibitem[Wheeler \& Akiyama(2004)]{wheeler04}Wheeler, J. C. \& Akiyama, S. 2004,
astro-ph/0412382
\bibitem[Woosley, Langer, \& Weaver(1993)]{woosley93}Woosley, S. E., Langer, N.,
\& Weaver, T. A. 1993, \apj, 411, 823
\bibitem[Yamasaki \& Yamada(2005)]{yamasaki05}Yamasaki, T., \& Yamada, S.
2005, \apj, 623, 1000
\bibitem[Young et al.(2005)]{young05}Young, P. A., Fryer, C. L., Hungerford, A.,
Arnett, D., Rockefeller, G., Timmes, F. X., Voit, B., Meakin, C., \& Eriksen,
K. A. 2005, astro-ph/0511806
\bibitem[Zatsarinny et al.(2004a)]{zatsarinny04a}Zatsarinny, O., Gorczyca, T.
W., Korista, K. T., Badnell, N. R., \& Savin. D. W. 2004, \aap, 417, 1173
\bibitem[Zatsarinny et al.(2003)]{zatsarinny03}Zatsarinny, O., Gorczyca, T. W.,
Korista, K. T., Badnell, N. R., \& Savin. D. W. 2003, \aap, 412, 587
\bibitem[Zatsarinny et al.(2004b)]{zatsarinny04b}Zatsarinny, O., Gorczyca, T.
W., Korista, K. T., Badnell, N. R., \& Savin. D. W. 2004, \aap, 426, 699
\bibitem[Zatsarinny et al.(2005)]{zatsarinny05}Zatsarinny, O., Gorczyca, T. W.,
Korista, K., Fu, J., Badnel, N. R., Mitthumsiri, W., \& Savin, D. W. 2005,
\aap, 438, 743
\bibitem[Zhang, Woosley, \& Heger(2004)]{zhang04}Zhang, W., Woosley, S. E.,
\& Heger, A. 2004, \apj, 608, 365
\bibitem[Zhekov et al.(2005)]{zhekov05}Zhekov, S. A., McCray, R. M., Borkowski,
K. J., Burrowns, D. N., \& Park, S. 2005, \apj, 628, L127
\end{thebibliography}
\end{document}